\documentclass[12pt,preprint]{aastex}
\usepackage{amsmath}
\usepackage{lineno}
\usepackage{color}

\def\jcap{JCAP}
\def\beq{\begin{equation}}
\def\eeq{\end{equation}}
\def\ben{\begin{eqnarray}}
\def\een{\end{eqnarray}}

\def\munit{M_{\odot}}
\def\ms{M_{\star}}

\def\mts{M_{\star, t2}}
\def\mtp{M_{\star, t1}}

\def\bjt{{\bf J}_{t}}
\def\bjb{{\bf J}_{b}}
\def\bjs{{\bf J}_{s}}
\def\ej{{\bf e}_{1}}
\def\ei{{\bf e}_{2}}
\def\en{{\bf e}_{3}}
\def\tlc{\Delta\tilde{\lambda}_{c}}
\def\cost{\cos\theta}

\begin{document}
\title{Detection of the Mass-Dependent Dual Type Transition of Galaxy Spins in IllustrisTNG Simulations}
\author{Jounghun Lee\altaffilmark{1}, Jun-Sung Moon\altaffilmark{2},
Suho Ryu\altaffilmark{1}, Suk-Jin Yoon\altaffilmark{2}}
\altaffiltext{1}{Astronomy Program, Department of Physics and Astronomy, 
Seoul National University, Seoul 08826, Republic of Korea 
\email{jounghun@astro.snu.ac.kr,shryu@astro.snu.ac.kr}}
\altaffiltext{2}{Department of Astronomy, Yonsei University, Seoul, 03722, Republic of Korea 
\email{moonjs@yonsei.ac.kr,sjyoon0691@yonsei.ac.kr}}

%%%%%%%%%%%%%%%%%%%%%%%%%%%%%%%%%%%%%%%%%%%%%%%%%%%%%%%%%%
\begin{abstract}
A numerical detection of the mass-dependent spin transition of the galaxies is presented. 
Analyzing a sample of the galaxies with stellar masses in the range of $10^{9}< (\ms/\munit)\le 10^{11}$
from the IllustrisTNG300-1 simulations, we explore the alignment tendency between the galaxy baryon spins and the three 
eigenvectors of the linearly reconstructed tidal field as a function of $\ms$ and its evolution in the redshift range of 
$0\le z \le 1.5$.  Detecting a significant signal of the occurrence of the mass-dependent transition of the galaxy spins, 
we show that the centrals differ from the satellites in their spin transition type. 
As $\ms$ increases beyond a certain threshold mass, the preferred directions of the central galaxy spins transit from 
the minor to the intermediate tidal eigenvectors (type two) at $z=0.5$ and $1$, while those of the satellites transit from 
the minor to the major tidal eigenvectors (type one) at $z=1$ and $1.5$. 
It is also shown that the mass range and type of the spin transition depend on the galaxy morphology, degree of 
the alignments between the baryon and total spin vectors as well as on the environmental density.  
 Meanwhile, the stellar spins of the galaxies are found to yield a weak signal of the T1 transitions at $z=0$, whose 
strength and trend depend on the degree of the alignments between the stellar and baryon spins.  The possible 
mechanisms responsible for the T1 and T2 spin transitions are discussed.
\end{abstract}
\keywords{Unified Astronomy Thesaurus concepts: Large-scale structure of the universe (902)}
%%%%%%%%%%%%%%%%%%%%%%%%%%%%%%%%%%%%%%%%%%%%%%%%%%%%%
\section{Introduction}\label{sec:intro}

The halo spin transition designates the phenomenon that the dark matter (DM) halos embedded in filaments prefer 
the directions parallel (perpendicular) to the filament axes in their spin orientations if their masses are below (above) 
a certain threshold mass, $M_{t}$ \citep[e.g.,][]{ara-etal07,hah-etal07b,paz-etal08,cod-etal12,tro-etal13,lib-etal13,AY14,
dub-etal14,for-etal14,cod-etal15a,cod-etal15b,cod-etal18,gan-etal18,wan-etal18,lee19,kra-etal20}.  
The spin transition phenomena were first detected at low redshifts $z<1$ in DM only N-body simulations, which determined 
$M_{t}$ to be around $10^{12}M_{\odot}$ \citep[e.g.,][]{ara-etal07,cod-etal12,tro-etal13,lib-etal13}. In the follow-up works, it was found 
that the occurrence of the halo spin transition was not confined to the filamentary environments but quite universally witnessed 
in the cosmic web, although the value of $M_{t}$ itself depends on the web type \citep[e.g.,][]{cod-etal15a,WK17,cod-etal18,kra-etal20,LL20}.
Since it is a manifestation of the difference between the low-mass and high-mass halos in their interactions with the cosmic web, the mass 
dependent spin transition is expected to shed a new light on the galaxy assembly bias \cite[e.g.,][]{cod-etal12,gan-etal18,kra-etal20,son-etal21}. 

Furthermore, it was also recently demonstrated by a N-body experiment that the value of $M_{t}$ sensitively depends on the dark 
energy equation of state and neutrinos mass \citep{lee-etal20,LL20}, which implied that the halo spin transition, if observed on the galaxy 
scale, may be in principle a powerful complementary near-field probe of the background cosmology. 
An observational detection of the halo spin transition, however, is often plagued by low accuracy involved in the determination of 
the galaxy spin directions. Moreover, the expected misalignment in the spin direction between the luminous galaxies and the 
underlying host DM halos \citep[e.g.,][]{hah-etal10,ten-etal14,vel-etal15,chi-etal17,cod-etal18} complicates a theoretical prediction for the 
galaxy spin transition trends, requiring a thorough hydrodynamical test of its occurrence to precede any observational detection.

Recent numerical studies based on hydrodynamical simulations, however, could not draw a conclusive answer to the vital question 
of whether or not the galaxy spin vectors also experience similar mass-dependent transitions. 
\citet{dub-etal14} utilized the galaxy sample from the Horizon-AGN hydrodynamical simulations and found that the signal of the spin transition 
is significant at high redshifts between $1.2< z < 1.8$. The results of \citet{cod-etal18} from the same hydrodynamical simulations indicated no 
significant signal of the galaxy spin transition at $z<0.5$ due to the rapid decrease of the strengths of the alignments between the galaxy 
spins and the filament axes with the decrement of redshifts. This result was confirmed by \citet{ballet2} who found the galaxy spins 
from the EAGLE hydrodynamic simulations \citep{eagle} to be consistently aligned with the directions perpendicular to the filament axes. 
These hydrodynamical results failed to match the recent observational evidence found by \citet{wel-etal20} for the occurrence of the 
mass-dependent galaxy spin transitions at low redshifts from the SAMI (Sydney-AAO Multi-object Integral Field 
Spectrograph) Galaxy Survey.

In contrast, \citet{wan-etal18} presented a detection of significant signal of the mass-dependent spin transition from the Illustris-1 simulations 
\citep{illustris-1} even at $z=0$. Their claim garnered a support from \citet{kra-etal20} who demonstrated that the galaxies from the SIMBA 
hydrodynamical simulations \citep{simba} indeed yielded a signal of the mass-dependent spin transition at $z=0$, although the 
signal is substantially weaker than those from the DM halos \citep{shi-etal21}. They explained that the inconsistencies among different 
hydrodynamical simulations on the significance of galaxy spin transition signal at low redshifts are likely caused by the differences in the scales 
and thicknesses of the identified filaments  (K. Kraljic in private communication). 

Meanwhile, \citet{LL20} developed a new algorithm for the spin transition threshold, in their attempt to universally describe its web-type 
dependence with the help of a N-body simulation \citep[see also][]{lee-etal20}.  Rather than measuring the spin directions of DM halos relative to the 
surrounding filament axes or sheet planes, they determined them relative to the three eigenvectors of the local tidal tensors, varying the smoothing 
scale. Finding that the preferred directions of the halo spins exhibit a mass-dependent transition from the  minor to the  intermediate
tidal eigenvectors at $z=0$, \citet{LL20} suggested that the spin transition threshold should be determined as the mass range at which the 
Kolmogorov--Smirnov (KS) test rejects the null hypothesis of $p(\cost_{2})=p(\cost_{3})$ at a confidence level lower than $99.9\%$. Here, $p(\cost_{2})$ and 
$p(\cost_{3})$ represent the probability densities of the cosines of the angles of the halo spin axes from the intermediate and minor tidal eigenvectors, respectively. 
This new methodology turned out to have several distinct advantages.  First, it can efficiently determine the transition threshold even when the size 
of a halo sample is quite small.  Second, it can sort out a false signal of the spin transition in case that the halo spin directions are aligned with both 
of the intermediate and minor tidal eigenvectors.  Third, it is free from the ambiguity associated with the fact that there is no established unique way to 
identify the filaments and sheets from the halo distributions. 
Fourth and most importantly, it can consistently and quantitatively describe how the transition trend varies with the scale of the cosmic web. 

In this Paper, we revisit the issue of the galaxy spin transition by applying the algorithm of \citet{LL20} to a high-resolution hydrodynamical 
simulation and attempt to answer the aforementioned vital question. The content of each Section can be outlined as follows.  
In Section \ref{sec:data} a short description of the numerical data and algorithm of \citet{LL20} is presented.  
In Section \ref{sec:main} the dependence of the galaxy spin transitions on the redshifts, environmental density, and 
galaxy properties are explored.
In Section \ref{sec:con} the key results are summarized and their implications are discussed. 

\section{Numerical Data and Analysis}\label{sec:data}

We utilize the data from the IllustrisTNG 300-1 hydrodynamical simulation \citep{tngintro1, tngintro2, tngintro3, tngintro4, tngintro5, illustris19}
conducted on a periodic box of linear size 
$302.6\,$Mpc for a Planck cosmology \citep{planck16}. The simulation accommodated various baryon physics such as star formation, 
stellar feedback outflow, radiative cooling and so forth  \citep{wei-etal17,pil-etal18} to track the evolution of $2500^{3}$ 
gas cells as well as of equal number of DM particles whose individual masses were set at $m_{b}=1.1\times 10^{7}\munit$ and 
at $m_{\rm dm}=5.9\times 10^{7}\munit$, respectively.
From the IllustrisTNG website\footnote{https://www.tng-project.org/data/}, we extract the catalogs of the bound halos and their subhalos that were 
identified via the friends-of-friends (FoF) and SUBFIND algorithms \citep{subfind}, respectively, at  four different redshfits, $z=0,\ 0.5,\ 1$ and 
$ 1.5$.  

The IllustrisTNG catalog lists sundry properties of each subhalo such as  the comoving position (${\bf x}_{c}$) of the center, peculiar velocity (${\bf v}_{c}$) 
of the center, total spin vector ($\bjt$) defined as the angular momentum of all (both DM and baryonic) member particles, and
stellar mass ($\ms$) within twice the stellar half-mass radius ($2R_{\star,1/2}$). Yet, it provides no information 
on the spin vector of  total baryonic particles ($\bjb$) belonging to each subhalo nor on the spin vector of stellar cells ($\bjs$) within $2R_{\star,1/2}$.
Using the available information on the member particles, however, we directly evaluate $\bjb$ of each subhalo as 
%%%%%%%%%%%%%%%%%%%%%%%%%%%%%%%%%%%%%%%%%%%%%%%%%%%%%
\begin{equation}
\label{eqn:gas_spin}
\bjb = \sum_{i=1}^{n_{\rm b}}{m}_{b,i}\,[({\bf x}_{b,i}-{\bf x}_{c})\times({\bf v}_{b,i}-{\bf v}_{c})]\, ,\\
\end{equation}
%%%%%%%%%%%%%%%%%%%%%%%%%%%%%%%%%%%%%%%%%%%%%%%%%%%%%
where $n_{\rm b}$ denotes the number of  all baryonic (both gas and star) particles belonging to each subhalo, and $m_{b,i}$, ${\bf x}_{b,i}$ and 
${\bf v}_{b,i}$ are the mass, comoving position and peculiar velocity of the $i$th baryonic particle, respectively. 
In a similar manner, we also evaluate $\bjs$. To minimize possible spurious signals caused by the inaccurate measurement of 
$\bjb$ and $\bjs$ \citep{bet-etal07}, we consider only those subhalos containing $100$ or more stellar particles within $2R_{\star,1/2}$ 
corresponding to $\ms\ge 1.1\times 10^{9}\munit$. In the third (from the left) column of Table \ref{tab:ng} is listed the numbers of the subhalos, 
$N_{g}$, with this stellar mass cut-off at each redshift.
 
To linearly construct the tidal fields from the simulation data, we first frame a dimensionless density contrast, 
$\delta_{r}({\bf x})$, on $256^{3}$ grid points with the help of the cloud-in-cell method applied to the FoF halo sample. 
Basically, counting the DM and baryon particles belonging to the FoF halos whose positions fall in each grid, we determine the mass densities, 
$\rho_{r}({\bf x})$ (total mass divided by the grid volume), via the cloud-in-cell interpolation. 
To take into account those particles which do not belong to the FoF halos, we make a simple assumption that they 
are distributed more or less uniformly over the grids and add their uniform density to $\rho_{r}({\bf x})$ as a background. 
This assumption has been justified in the previous works  which found that the eigenvectors of the tidal fields 
determined from the FoF halos under this assumption of uniform distribution of non-FoF particles were well aligned 
with those rigorously determined from the particle data \citep{lee-etal20,LL20}.

The dimensionless density contrast field at each grid is evaluated as the residual overdensity,  
$\delta_{r}({\bf x}) \equiv [\rho_{r}({\bf x}) - \langle\rho_{r}\rangle]/\langle\rho_{r}\rangle$ where $\langle\rho_{r}\rangle$ is the average 
including the background uniform density taken over $256^{3}$ grid points. The second (from the left) columns of 
 Table \ref{tab:ng} list the numbers of the FoF halos,  $N_{\rm fof}$, used for the construction of  $\delta_{r}({\bf x})$
at each redshift. 
The tidal field, ${\bf T}=(T_{ij})$, smoothed on the typical cluster scale of $R_{f}=2\,$Mpc is then obtained 
by following the same two-step procedure as described in \citet{lee-etal20}: The Fourier-transformation of $\delta_{r}({\bf x})$ into $\tilde{\delta_{r}}({\bf k})$ 
and then the inverse Fourier transformation of $\hat{k}_{i}\hat{k}_{j}\tilde{\delta}({\bf k})\exp\left(-k^{2}R^{2}_{f}/2\right)$. 
It may be worth here explaining the logic behind our initial setting $R_{f}=2\,$Mpc. The majority of the galaxies are embedded in the galaxy clusters whose 
typical size is $\sim 2\,$Mpc.   On the smaller scales than $2\,$Mpc, the linearly reconstructed tidal field would no longer be a valid approximation, while 
on the larger scales, the tidal fields are expected to be less strongly aligned with the subhalo spin vectors \citep{lee19,shi-etal21}.   

\begin{figure}[ht]
\begin{center}
\plotone{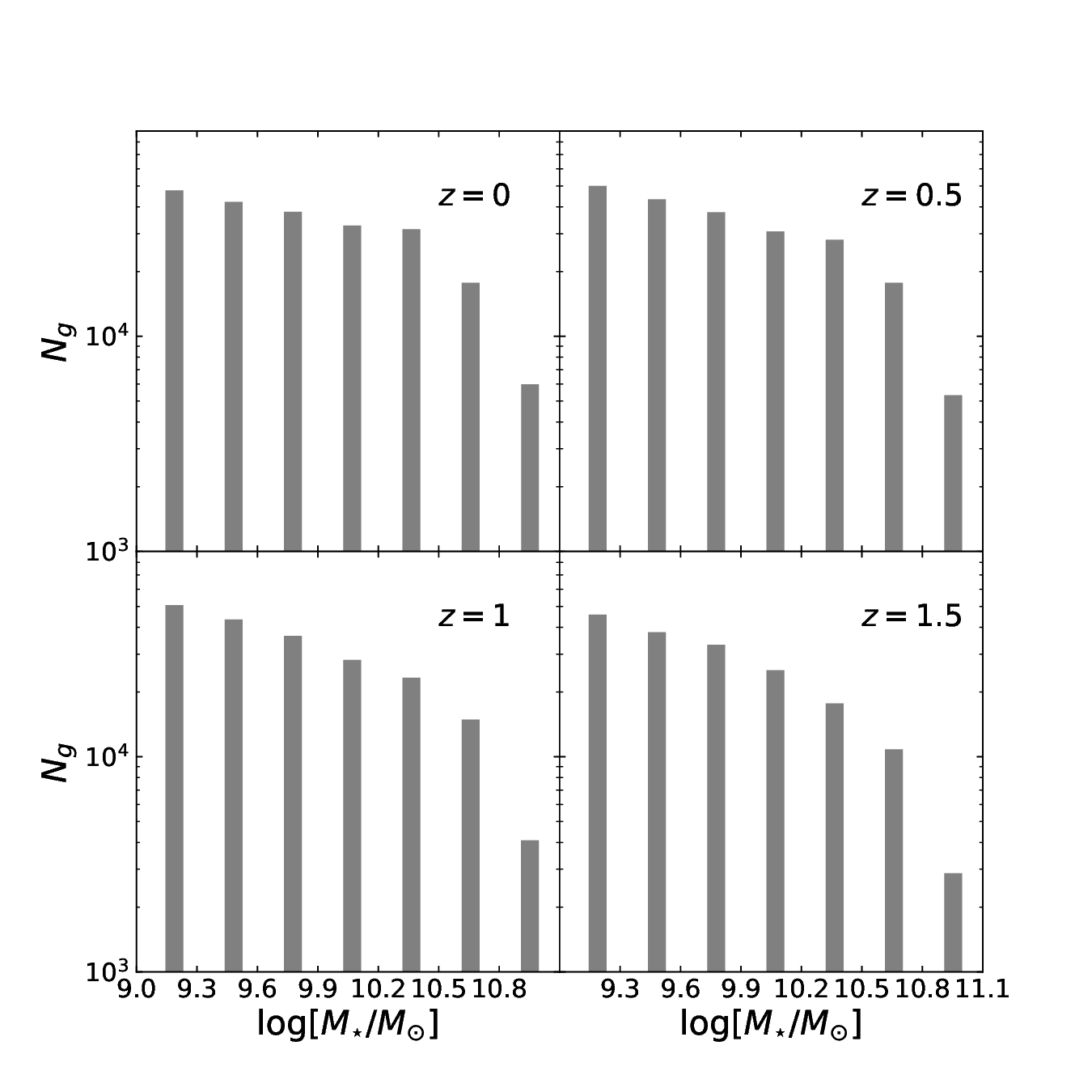}
\caption{Number distributions of the galaxies versus $\log(\ms/\munit)$ at  four different redshifts.}
\label{fig:mdis}
\end{center}
\end{figure}

Locating the grid in which the position of each subhalo falls, we diagonalize $(T_{ij})$ at the grid to find its three eigenvalues, 
$\{\lambda_{1}, \lambda_{2},\lambda_{3}\}$ in a decreasing order and the corresponding major, intermediate minor eigenvectors 
$\{\ej,\ei,\en\}$ as well. 
For each galaxy, we calculate $\cost_{b,i}\equiv \vert\bjb\cdot{\bf e}_{i}\vert/\vert\bjb\vert$ for $i\in \{1,2,3\}$ to quantify 
their spin alignments with respect to the tidal eigenvectors. The closer to the unity (zero) the value of $\cost_{b,i}$ is, the more 
strongly the spin vector is aligned (anti-aligned) with the tidal eigenvector ${\bf e}_{i}$. 
Splitting the range of $\log\ms$ into several short bins of equal length, we count the numbers of the subhalos, $N_{g}$, 
which fall in each bin ( Figure \ref{fig:mdis}). Then, we evaluate the ensemble average and its associated error for each $i\in \{1,2,3\}$ as
%%%%%%%%%%%%%%%%%%%%%%%%%%%%%%%%%%%%%%%%%%%%%%%%%%%%%%%%%%%%%%%%%%%%%%%
\begin{eqnarray}
\label{eqn:mean} 
\langle\cost_{b,i}\rangle &=& N^{-1}_{g}\sum_{i=1}^{N_{g}} \cost_{b,i}\, , \\
\sigma_{\cost_{b,i}} &=& \left[N^{-1}_{g}\sum_{i=1}^{N_{g}} \left(\cost_{b,i}-\langle\cost_{b,i}\rangle\right)^{2}\right]^{1/2}\, .
\end{eqnarray}
%%%%%%%%%%%%%%%%%%%%%%%%%%%%%%%%%%%%%%%%%%%%%%%%%%%%%%%%%%%%%%%%%%%%%%%%%
In a similar manner, we also evaluate $\langle\cost_{t,i}\rangle$ with $\sigma_{\cost_{t,i}}$ and 
$\langle\cost_{s,i}\rangle$ with $\sigma_{\cost_{s,i}}$ for each $i\in \{1,2,3\}$. 

\section{Dual-Type Transitions of the Galaxy Baryon Spins}\label{sec:main}

\subsection{Variation with redshifts and environments}\label{sec:var}

Figure \ref{fig:ez} plots $\langle\cost_{b,i}\rangle$ (left panel), $\langle\cost_{s,i}\rangle$ (middle panel) and $\langle\cost_{t,i}\rangle$ (right panel) 
for $i\in \{1,2,3\}$ in the mass range of $9<\log(\ms/\munit)\le 11.1$ at  $z=0,\ 0.5,\ 1$ and $1.5$.  
The results in the higher mass-ranges $\log(\ms/\munit)> 11.1$ are left out due to their larger errors.  
Hereafter, the subhalos with $9<\log(\ms/\munit)\le 11.1$ are referred to as the galaxies whose number distributions $N_{g}$ are plotted in Figure \ref{fig:mdis}. 
As can be seen, the alignment tendencies of the baryon spins are similar to those of the total spins,  while the stellar spins show a distinctively unique  
tendency of being dominantly aligned with $\ej$ in the high mass section at $z=0$.   
Despite that they show such conspicuously different alignment tendencies and true as it is that they are more readily observable in practice,  the stellar spins 
are not adequate for the study of the galaxy spin transition phenomena since $\{\langle\cost_{s,i}\rangle\vert i=1,2,3\}$ has the lowest amplitudes in the whole mass 
range, showing the most noisy variations with $\log\ms$.  
Meanwhile, we note that $\{\langle\cost_{b,i}\rangle\vert i=1,2,3\}$ exhibits the highest amplitudes and the least noisy variations with $\log\ms$ at all of the 
 four redshifts.  Hereafter, for the investigation of the mass-dependent spin transition of the galaxies, we will mainly focus on the 
baryon spins, calling them the galaxy spins unless otherwise stated. 

Before commencing this investigation, however, it should be worth mentioning here that the expression, "the alignments between the galaxy spins and tidal 
eigenvectors", actually do not refer to the extreme case that their directions are perfectly parallel to each other. 
Rather, this expression refers to the tendencies of the galaxy spin orientations not to be random (i.e., to be anisotropic) in the tidal principal frame. 
In other words, when it is conventionally and somewhat misleadingly expressed that the galaxy spin vectors are {\it aligned} with one of the tidal eigenvectors, 
it actually corresponds to the case that the probability density distributions of the cosines of the angles between the spin vectors and the tidal eigenvectors 
show statistically significant deviations from the uniform distribution \citep{LP00,LP01}. 

Let us classify the galaxy spin transition into two different types. If the preferred directions of the galaxy spins undergo a mass-dependent transition between 
$\ej$ and $\en$ ($\ei$ and $\en$), it is classified as the type one (type two) spin transition denoted by T1 (T2). 
To determine where this transition occurs, we adopt the rigorous procedure proposed by \citet{LL20} based in the Kolmogorov--Smirnov (KS) test.  
Let $p(\cost_{b,i})$ denote the probability density distribution of $\cost_{b,i}$ and define its cumulative distribution as
$P(<\cost_{b,i})\equiv \int^{\cost_{b,i}}_{0}p(\cost^{\prime}_{b,i})d\cost^{\prime}_{b,i}$ for $i\in \{1,2,3\}$.  Set up a null hypothesis 
that two random variables, $\cost_{b,3}$ and $\cost_{b,j}$, are independently identically distributed at a given mass bin, 
[$p(\cost_{b,j})\overset{\text{i.i.d.}}{\sim} p(\cost_{b,3})$] for $j\in \{1,2\}$. 

Perform a KS test of this hypothesis by calculating the following KS statistics at each mass bin,  
%%%%%%%%%%%%%%%%%%%%%%%%%%%%%%%%%%%%%%%%%%%%%%%%%%%%%%%%%%%%%%%%%%
\begin{eqnarray}
\label{eqn:ks}
\tilde{D}_{\rm max, t1}(\log\ms) &\equiv& \frac{\sqrt{N_{g}}}{2}{\rm max}\vert P(<\cost_{b,1}) - P(<\cost_{b,3})\vert\, , \\
\tilde{D}_{\rm max, t2}(\log\ms) &\equiv& \frac{\sqrt{N_{g}}}{2}{\rm max}\vert P(<\cost_{b,2}) - P(<\cost_{b,3})\vert\, , 
\end{eqnarray}
%%%%%%%%%%%%%%%%%%%%%%%%%%%%%%%%%%%%%%%%%%%%%%%%%%%%%%%%%%%%%%%%%%%%%%%%
which are the weighted measures of the maximum differences between $P(<\cost_{b,1})$ and $P(<\cost_{b,3})$ and between $P(<\cost_{b,2})$ and 
$P(<\cost_{b,3})$, respectively. If $\tilde{D}_{\rm max, tj}(\log\ms)\ge 1.949$ at a given mass bin for $j\in \{1,2\}$, then the null hypothesis is rejected 
by the KS test at the $99.9\%$ confidence level there. 

The T1 (T2) spin transition zones correspond to the mass bins where the condition of $\tilde{D}_{\rm max, t1}(\log\ms)<1.949$ [$\tilde{D}_{\rm max, t2}(\log\ms)<1.949$] 
is so satisfied that the null hypothesis is not rejected at the $99.9\%$ confidence level. Although it was not explicitly stated in \citet{LL20},  the T1 (T2) 
spin transition zone must satisfy an additional condition, $\tilde{D}_{\rm max, t1}(\log\ms)>1.949$ [$\tilde{D}_{\rm max, t2}(\log\ms)>1.949$] in the adjacent 
mass bins. This additional condition is required to guarantee that the KS statistics reach the local minima in the spin transition zones.

The left panels of Figure \ref{fig:clz} show $\tilde{D}_{\rm max, t2}(\log\ms)$ (red lines) at  $z=0,\ 0.5,\ 1$ and $1.5$.  
In each panel the horizontal dotted line corresponds to the critical value of $\tilde{D}_{c}=1.949$ corresponding to the $99.9\%$ confidence level 
in the KS test.  The mass ranges of the T2 spin transition zones (denoted by $\mts$) are listed in the first four rows of Table \ref{tab:tzone1}. 
Note that clear signals of the T2 spin transition are found at $z=0,\ 0.5$ and $1$ but not at $z=1.5$ and that $\mts$ falls in the higher mass range at 
$z=0$ than at $z=1$.  Despite that no significant signal of the the T2 spin transition zone is found at $z=1.5$, both of $\langle\cost_{b,3}\rangle$ and 
$\langle\cost_{b,2}\rangle$ at $z=1.5$ change in a mass dependent way, becoming $\langle\cost_{b,3}\rangle=\langle\cost_{b,2}\rangle$ at the smallest 
mass bin of $9.0<\log\ms/\munit\le 9.3$, as shown in Figure \ref{fig:ez}. We also compute $\tilde{D}_{\rm max, t1}(\log\ms)$ but find that 
at none of the mass bins the condition for the T1 spin transition is satisfied, which indicates no occurrence of the T1 spin transition.

To investigate how the variation of $R_{f}$ affects the  T2 spin transition zones, we repeat the whole analysis but for two different cases of 
$R_{f}=3\,$Mpc and $5\,$Mpc. The resulting $\{\langle\cost_{b,i}\rangle\vert i=1,2,3\}$ and $\tilde{D}_{\rm max, t2}(\log\ms)$ are plotted in 
Figure \ref{fig:ez_rf} and in the left panel of Figure \ref{fig:clz} (blue and green lines), respectively, while the resulting mass ranges of $\mts$ 
are listed in the second and third  four rows of Table \ref{tab:tzone1}. 
As can be seen,  the increment of $R_{f}$ slightly decreases the amplitudes of $\{\langle\cost_{b,i}\rangle\vert i=1,2,3\}$ at all of the four redshifts. 
At $z=1.5$ the T2 spin transition zone is found only for the case of $R_{f}=5\,$Mpc in the range of $9.3\le \log(\ms/M_{\odot})\le 10.2$. 
At $0\le z\le 1$, although the overall alignment trend does not change sensitively with $R_{f}$, the T2 spin transition zone tends to be substantially 
shifted to the higher mass range as $R_{f}$ increases. 
The increment of $R_{f}$ from $2$ to $5\,$Mpc induces approximately $150\%$ fractional change of the mean value of $\mts$. Hereafter, we will set 
the value of $R_{f}$ at $2\,$Mpc, for which case $\{\langle\cost_{b,i}\rangle\vert i=1,2,3\}$ exhibit the highest amplitudes.

If the galaxies are located in the regions where any two or all three of the tidal eigenvalues are very similar to one another (e.g., prolate/oblate or 
spherical regions), then the determinations of the corresponding tidal eigenvectors could be contaminated by the eigenvalue degeneracy, 
which in consequence could also deter the proper measurements of the spin transition zones. To avoid this possible contamination caused by 
the eigenvalue degeneracy, we calculate the differences among the rescaled tidal eigenvalues as 
%%%%%%%%%%%%%%%%%%%%%%%%%%%%%%%%%%%%%%%%%%%%%%%%%%%%%%%%%%%%%%%%%%
\begin{equation}
\label{eqn:eigr}
\Delta\tilde{\lambda}_{12} \equiv \frac{\vert\lambda_{1}-\lambda_{2}\vert}{\sqrt{\lambda^{2}_{1}+\lambda^{2}_{2}+\lambda^{2}_{3}}}\, ,\quad
\Delta\tilde{\lambda}_{23} \equiv \frac{\vert\lambda_{2}-\lambda_{3}\vert}{\sqrt{\lambda^{2}_{1}+\lambda^{2}_{2}+\lambda^{2}_{3}}}\, , 
\end{equation}
%%%%%%%%%%%%%%%%%%%%%%%%%%%%%%%%%%%%%%%%%%%%%%%%%%%%%%%%%%%%%%%%%%%%%%%%
and select only those galaxies located in the regions where two conditions of 
$\Delta\tilde{\lambda}_{12}>\Delta\tilde{\lambda}_{c}$ and $\Delta\tilde{\lambda}_{12}>\Delta\tilde{\lambda}_{c}$ are simultaneously satisfied, 
where $\Delta\tilde{\lambda}_{c}$ is a certain lower limit.

Figure \ref{fig:eigr} plots the distributions of $\Delta\tilde{\lambda}_{12}$ (left panels) and $\Delta\tilde{\lambda}_{23}$ (right panels) at 
the four redshifts.  Considering two different cases of $\Delta\tilde{\lambda}_{c}=0.1$ and $0.2$, and setting $R_{f}=2\,$Mpc,  
we repeat the whole analysis with the selected galaxies, the results of which are shown in Figure \ref{fig:ez_eig} and in the right panels 
of Figure \ref{fig:clz}. 
The first and second  four rows of Table \ref{tab:tzone2} list the mass ranges of $\mts$ for the cases of $\Delta\tilde{\lambda}_{c}=0.1$ and 
$\Delta\tilde{\lambda}_{c}=0.2$, respectively.  As can be seen, the variation of $\Delta\tilde{\lambda}$ from $0.0$ to $0.2$ leads $\mts$ to 
be slightly shifted to a lower mass range at $z=0$ and $0.5$.  At $z=1$, no spin transition zone is found for the case of 
$\Delta\tilde{\lambda}_{c}=0.2$, while the other two cases yield the same spin transition zone. 
The behavior of $\tilde{D}_{\rm max}$ for the case of $\Delta\tilde{\lambda}_{c}=0.2$, however, is quite similar to those for the other two cases, dropping 
below the threshold value, $1.949$, in the same mass range of $9.3\le \log(\ms/M_{\odot})\le 9.9$. At $z=1.5$, no spin transition zone is found for 
all of the three cases. This result indicates  that the eigenvalue degeneracy does not severely contaminate the determination of the T2 spin transition zones. 
We also examine if the variations of $R_{f}$ and $\Delta\tilde{\lambda}_{c}$ can lead the T1 spin transition to occur but fail to find any signal.

To investigate if and how the spin transition zones depend on the {\it smoothed} density contrast, $\delta$, which is equal to $\sum_{i=1}^{3}\lambda_{i}$, 
we split the galaxy sample into two subsamples each containing the galaxies embedded in the high-density ($\delta > 1$) and in the low-density 
($\delta\le 1$) environments, respectively. 
Then, we separately determine $\{\langle\cost_{b,i}\rangle\vert i=1,2,3\}$ and $\tilde{D}_{\rm max, t2}$ for each subsample, 
the results of which are shown in Figure \ref{fig:ez_den}-\ref{fig:clz_den} and in the third and fourth four rows of Table \ref{tab:tzone2}. 
As can be seen, for the case of $\delta\le 1$, a signal of the T2 spin transition is found only at $z=0$ but not at $z> 0$. Whereas, for the case of 
$\delta>1$, the T2 spin transition signals are found at all of the four redshifts, showing a trend that $\mts$ falls in the lower-mass range 
at higher redshifts.  
It should be also noted that $\mts$ at $z=0$ for the case of $\delta\le 1$ falls in the lower-mass range than that for the case of $\delta>1$. 

\subsection{Dependence on the galaxy properties}\label{sec:dep}

To investigate the dependence of the spin transition trend on the galaxy morphology, we first determine the kinetic energy contributed 
by the corotational motion, $K_{\rm rot}$, for each galaxy composed of $n_{sp}$ stellar particles within $2R_{\star,1/2}$ such that 
%%%%%%%%%%%%%%%%%%%%%%%%%%%%%%%%%%%%%%%%%%%%%%%%%%%%%%%%%%%%%%%%%%
\begin{equation}
\label{eqn:rot}
K_{\rm rot} = \sum_{i=1}^{n_{sp}}\frac{1}{2}\,{m}_{p,i}\,\left(\frac{j_{s,i}}{{m}_{p,i}d_{i}}\right)^2\, ,
\end{equation}
%%%%%%%%%%%%%%%%%%%%%%%%%%%%%%%%%%%%%%%%%%%%%%%%%%%%%%%%%%%%%%%%%%
where ${m}_{p,i}$, $j_{s,i}$ and $d_{i}$ denote the mass and the angular momentum along the direction of $\bjs$ and the projected distance to the 
central axis of the $i$th stellar particle within $2R_{\star,1/2}$, respectively. 
Taking its ratio to the total kinetic energy of the stellar particles as $\kappa_{\rm rot} \equiv K_{\rm rot}/K$, we classify the galaxies with 
$\kappa_{\rm rot} < 0.5$ ($\kappa_{\rm rot}\ge 0.5$) as spheroidals (disks) \citep{sal-etal12,cor-etal17,rod-etal17}. 

Given that the galaxy morphology is well known to be strongly correlated with the environmental densities, 
which in turn affect the spin transition zones as shown in Figures \ref{fig:ez_den}-\ref{fig:clz_den}, the effect of the morphology-density correlations on the 
spin transition zones should be first nullified to single out the morphology dependence of the spin transition zone. 
Binning the ranges of $\delta$ and selecting equal numbers of the spheroidals and disks at each $\delta$-bin, we create two controlled subsamples of the 
spheroidal and disk galaxies which share the identical $\delta$ distributions. 

We separately determine $\{\langle\cost_{b,i}\rangle\vert i=1,2,3\}$ and $\tilde{D}_{\rm max, t2}$ for each controlled sample, and show the results 
in Figures \ref{fig:ez_shape}-\ref{fig:clz_shape}.  
As can be seen,  we find a signal of the T2 spin transition only from the disks at $z=0.5$ and $1$, noting the same trend of $\mts$ falling 
in the lower-mass ranges at higher redshifts.  Table \ref{tab:tzone3} lists the corresponding ranges of $\mts$ in the first (from the top) two rows. 
As for the spheroidals, although they exhibit mass-dependent changes of the amplitudes of $\langle\cost_{2}\rangle$ and $\langle\cost_{3}\rangle$, 
no significant signal of the T2 spin transition is found in the whole mass range at all of the four redshifts.  

We also classify the galaxies into the centrals and satellites according to their subhalo total masses ($M_{s}$) and number of the subhalos that their host halos 
have ($N_{s}$). The centrals satisfy the condition of $M_{s}=M_{s,{\rm max}}\equiv {\rm max}\{M_{s,i}\vert 1\le i\le N_{s}\}$, while 
the condition for the satellites is $\{N_{s}>1$, $M_{s}< M_{s,{\rm max}}\}$. We control the subsamples of the centrals and satellites to have identical 
$\delta$-distributions at each redshift. 
Then, we separately determine $\{\langle\cost_{b,i}\rangle\vert i=1,2,3\}$, $\tilde{D}_{\rm max, t1}$ and $\tilde{D}_{\rm max, t2}$ from the controlled 
subsamples at the four redshifts, the results of which are shown in Figures \ref{fig:ez_status}-\ref{fig:clz_status} and in the second (from the top) 
five rows of Table \ref{tab:tzone3}. 
As can be seen, the spin transition type differs between the centrals and the satellites.  The former yields clear signals of the T2 spin transition at 
$z=0,\ 0.5$ and $1$,  while weak but significant signals of the T1 spin transition are found from the satellites at $z=1$ and $1.5$.  
As for the satellites at $z=0$ and $0.5$, although similar mass-dependent amplitude changes of $\langle\cost_{b,1}\rangle$ and $\langle\cost_{b,3}\rangle$ 
are found, their behaviors are too noisy for the KS test-based routine to locate the spin transition zones. As for the centrals at $z=1.5$, although they exhibit 
similar amplitudes and mass-dependent variations of $\langle\cost_{b,2}\rangle$ and $\langle\cost_{b,3}\rangle$, the larger errors fail us to locate the spin 
transition zones. 

To see if and how the alignments between the baryon and total spins are linked with the T1 and/or T2 spin transitions, 
we calculate the cosine of the angle between them,  $\cos\alpha\equiv \vert\bjb\cdot\bjt\vert$, for each galaxy and take its average 
over all of the galaxies at each redshift.  Splitting the galaxies into two subsamples satisfying the conditions of 
$\cos\alpha>\langle\cos\alpha\rangle$ and $\cos\alpha\le\langle\cos\alpha\rangle$, respectively, we control the two subsamples to 
have identical $\delta$-distributions. 
The resulting $\{\langle\cost_{b,i}\rangle\vert i=1,2,3\}$, $\tilde{D}_{\rm max, t1}$ and $\tilde{D}_{\rm max, t2}$ are displayed 
in Figures \ref{fig:ez_cosa}-\ref{fig:clz_cosa}. The corresponding spin transition zones are listed in the third (from the top) four rows in 
Table \ref{tab:tzone3}. 

As can be seen, clear signals of the mass-dependent T1 spin transition are found from those galaxies whose baryon spins are less strongly aligned 
with their total spins ($\cos\alpha\le \langle\cos\alpha\rangle$) at all of the four redshifts. Note that $\mtp$ falls in the higher mass range at higher 
redshifts, which trend is opposite to that of $\mts$. 
This result implies that the occurrence of the T1 spin transition, which is never found from the total spins, 
should be closely related to the mechanism responsible for the deviationg the directions of the baryon spins from those of the total spins. 
As for the galaxies whose baryon spins are more strongly aligned with their total spins ($\cos\alpha>\langle\cos\alpha\rangle$),  their spin alignment 
tendencies are quite similar to those of the spheroidal galaxies shown in Figure \ref{fig:ez_shape},  yielding no clear signal of the T2 spin transition 
but showing similar increment of $\langle\cost_{b,2}\rangle$ and decrement of $\langle\cost_{b,3}\rangle$ as $\log(\ms/\munit)$ becomes 
larger. 

\subsection{T1 transitions of the stellar spins}\label{sec:star}

In the light of the results shown in Figures \ref{fig:ez_cosa}-\ref{fig:clz_cosa}, we would like to investigate if and how the alignments of the galaxy 
{\it stellar} spins with the tidal eigenvectors, $\{\langle\cost_{s,i}\rangle\vert i=1,2,3\}$, depend on the alignments between the galaxy stellar and baryon 
spins, $\cos\beta\equiv \vert\bjs\cdot\bjb\vert$. We take the average, $\langle\cos\beta\rangle$, over all of the galaxies at each redshift and 
divide the galaxies into two subsamples each of which includes the galaxies with $\cos\beta>\langle\cos\beta\rangle$ and 
$\cos\beta\le \langle\cos\beta\rangle$, respectively. Controlling the two subsamples to have identical $\delta$-distributions as done 
in Section \ref{sec:dep}, we determine separately $\{\langle\cost_{s,i}\rangle\vert i=1,2,3\}$ and $\tilde{D}_{\rm max, t1}$ for each of the controlled 
subsamples. 

 Figure \ref{fig:esz_cosb} plots $\{\langle\cost_{s,i}\rangle\vert i=1,2,3\}$ at $z=0,\ 0.5,\ 1$ and $1.5$ from the samples with 
$\cos\beta\le \langle\cos\beta\rangle$ (left panels) and with $\cos\beta>\langle\cos\beta\rangle$ (right panels), revealing that the two subsamples yield 
very different alignment tendencies. 
From the galaxies whose stellar spins are less strongly aligned with the baryon spins ($\cos\beta\le \langle\cos\beta\rangle$), 
we find no signals of the mass dependent spin transitions but only consistent $\bjs$-$\ej$ alignments at all of the four redshifts. 
The overall amplitudes of $\langle\cost_{s,1}\rangle$ seem to decrease as $\log\ms$ decreases and $z$ increases. 
Meanwhile, from the galaxies whose stellar spins are more strongly aligned with the baryon spins ($\cos\beta>\langle\cos\beta\rangle$), 
we find weak but significant signals of the mass-dependent T1 spin transitions at $z=0$.  
At $z=0.5,\ 1$ and $1.5$, they exhibit mass-dependent changes of the amplitudes of  $\langle\cost_{s,2}\rangle$ and 
$\langle\cost_{s,3}\rangle$, which lead to $\langle\cost_{s,2}\rangle\sim \langle\cost_{s,3}\rangle$ at  $\log\left(\ms/\munit\right)\le 9.6$. 

Figure \ref{fig:clsz_cosb} plots $\tilde{D}_{\rm max, t1}$ at $z=0$ from the stellar spins of all galaxies (top panel) and from those with 
$\cos\beta>\langle\cos\beta\rangle$ (bottom panel), and Table \ref{tab:tzone4} lists the mass ranges of $\mtp$ for the two cases. 
Repeating the same analysis for two different cases of $\tlc=0.1$ and $0.2$, we have confirmed the robustness of this result against the 
eigenvalue degeneracy.  
Given the results shown in Figures \ref{fig:esz_cosb}-\ref{fig:clsz_cosb} and Table \ref{tab:tzone4}, we presume that the distinctively unique 
$\bjs$-$\ej$ alignment tendency exhibited by the stellar spins should be closely linked with the mechanism responsible 
for the misalignments between the stellar and baryon spins and that this mechanism should be more effective at lower redshifts. 

\section{Summary and Conclusion}\label{sec:con}

Analyzing the halo and subhalo catalogs from the IllustrisTNG 300-1 simulations \citep{illustris19} at $0\le z \le 1.5$, 
we have explored if and how the preferred spin directions of the galaxies with $9< \log(\ms/\munit)\le 11.1$ 
undergo any mass-dependent transitions in the principal frame spanned by the three eigenvectors of the local tidal fields smoothed 
on the scale of $R_{f}=2\,$Mpc. The diagnostics proposed by \citet{LL20} based on the KS test  has been employed to rigorously determine 
the range of the galaxy stellar mass where the spin transition occurs (spin transition zone) at each redshift.  
We have also investigated whether or not the occurrence of the spin transition depends on the environmental 
density contrast ($\delta$) as well as such galaxy properties as their morphology, central/satellite category, alignment between their baryon 
and total spins ($\cos\alpha$), and alignment between their stellar and baryon spins ($\cos\beta$). 

\begin{figure}[ht]
\begin{center}
\plotone{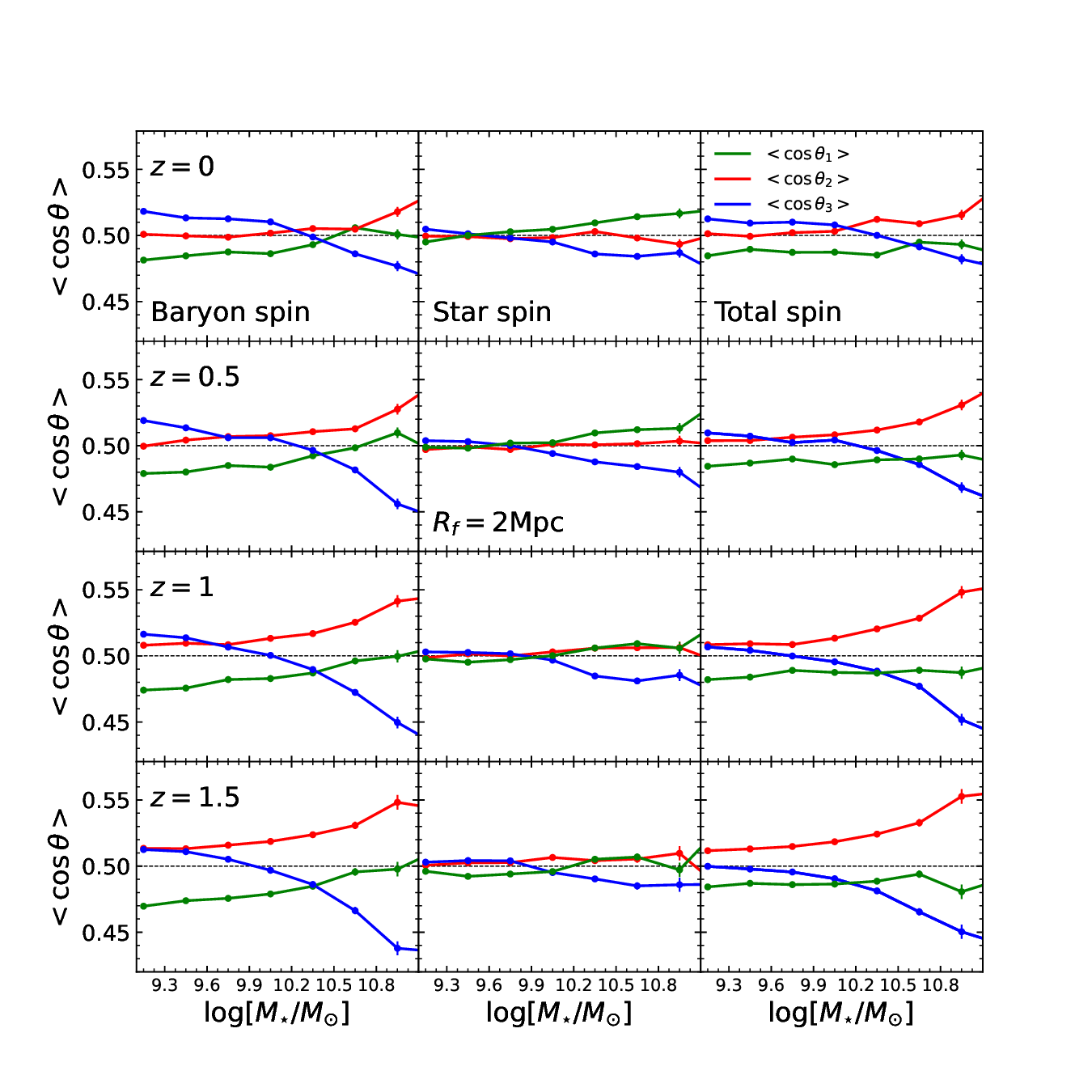}
\caption{Mean values of the cosines of the angles of the baryon, star and total spin vectors of the galaxies (left, middle and right panels, 
respectively) with the major, intermediate and minor eigenvectors of the tidal field (red, blue and green lines, respectively) smoothed 
on the scale of $2\,$Mpc at  four different redshfits. }
\label{fig:ez}
\end{center}
\end{figure}

\begin{figure}[ht]
\begin{center}
\plotone{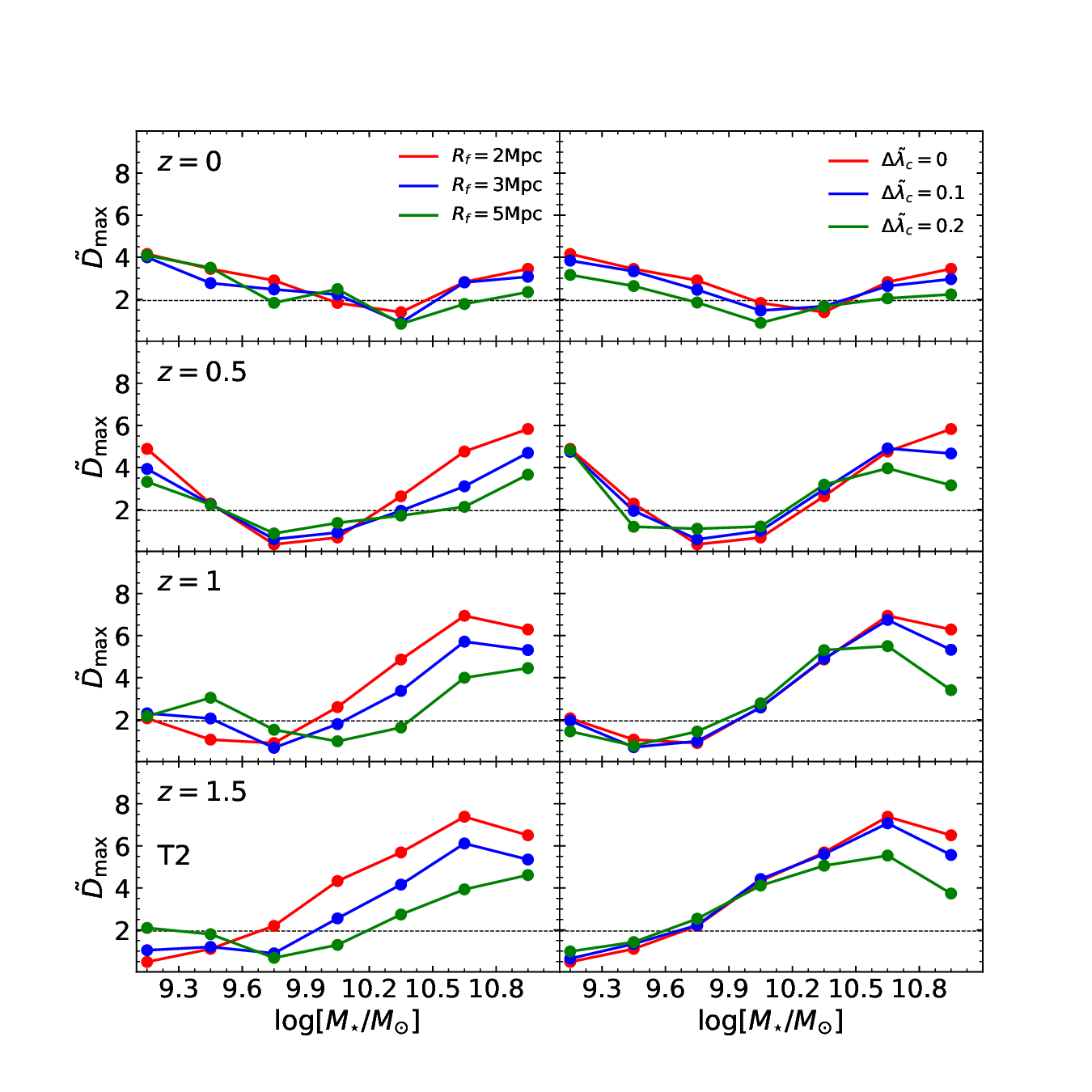}
\caption{KS statistics  for the determination of the T2 spin transition zone at four different redshfits, for three different 
cases of the smoothing scale $R_{f}$ (left panels) with the threshold for the rescaled eigenvalue difference, $\tlc$, set at $0$, 
and for three different cases of $\tlc$ (right panels) with $R_{f}$ set at $2\,$Mpc.}
\label{fig:clz}
\end{center}
\end{figure}

\begin{figure}[ht]
\begin{center}
\plotone{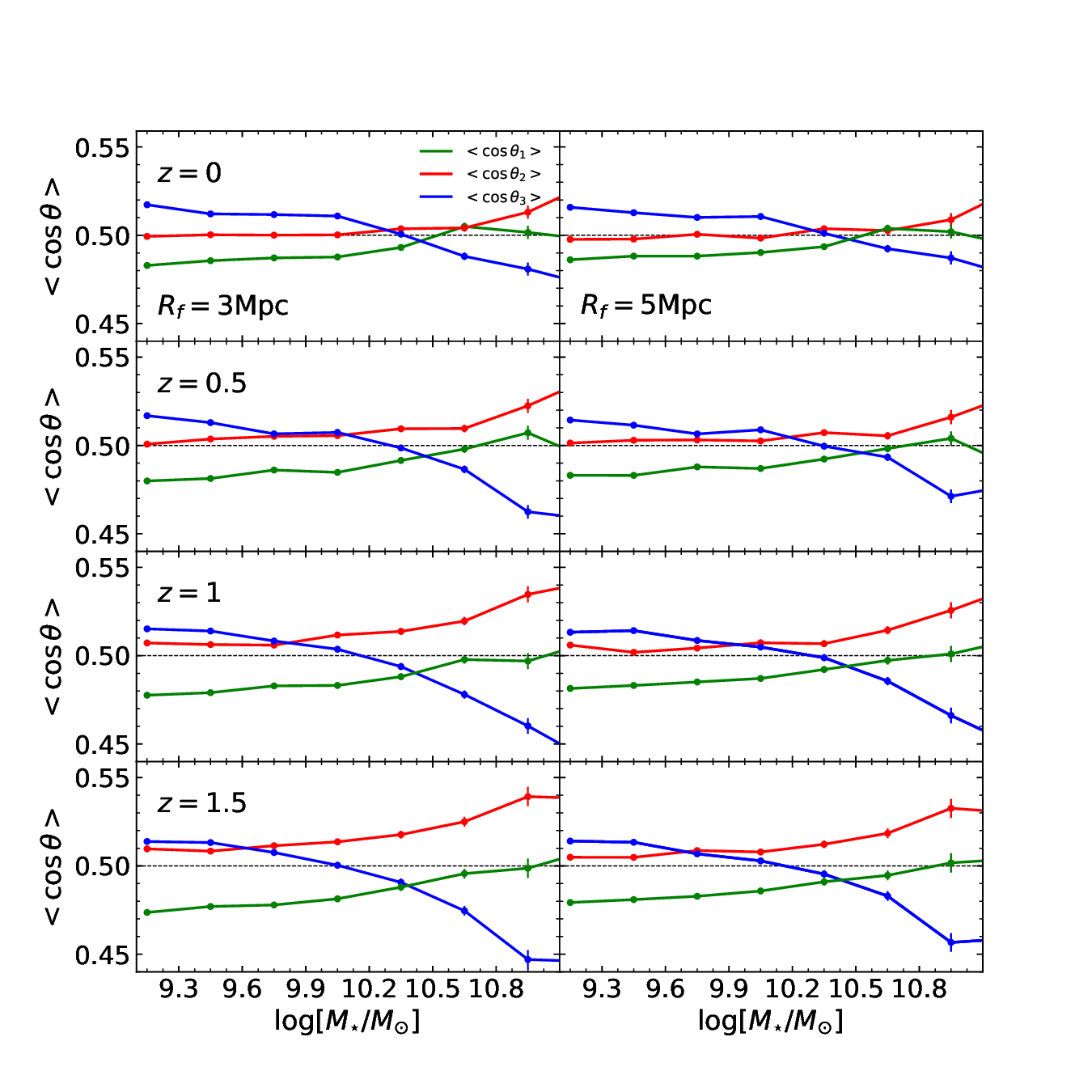}
\caption{Same as Figure \ref{fig:ez} but for two different cases of $R_{f}$.}
\label{fig:ez_rf}
\end{center}
\end{figure}

\begin{figure}[ht]
\begin{center}
\plotone{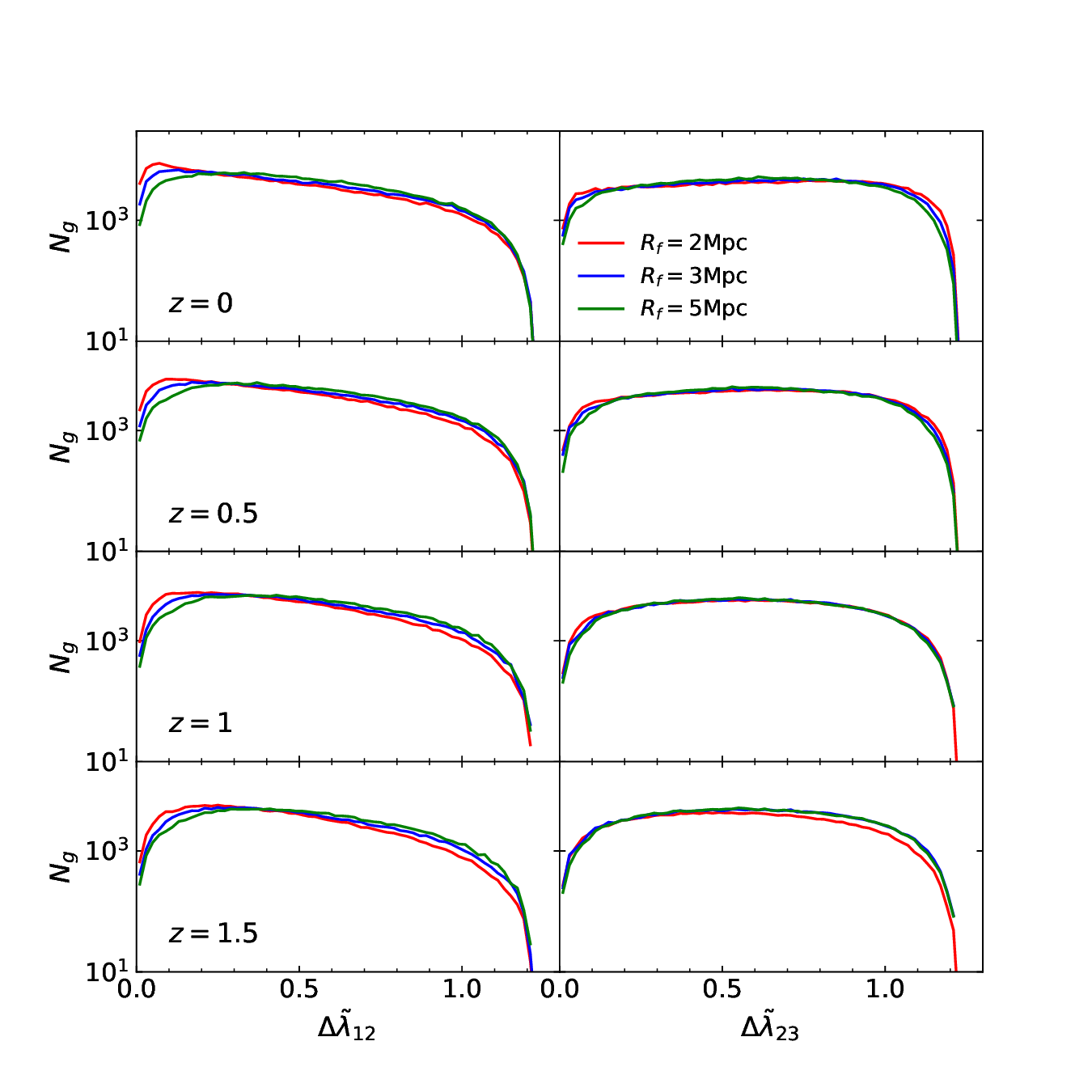}
\caption{Number distributions of the galaxies  versus the difference between the rescaled major and intermediate 
eigenvalues (left panels), and versus the difference between the rescaled intermediate and minor eigenvalues (right panels), 
defined in Equation (\ref{eqn:eigr}) with $R_{f}=2\,$Mpc at four different redshifts.}
\label{fig:eigr}
\end{center}
\end{figure}

\begin{figure}[ht]
\begin{center}
\plotone{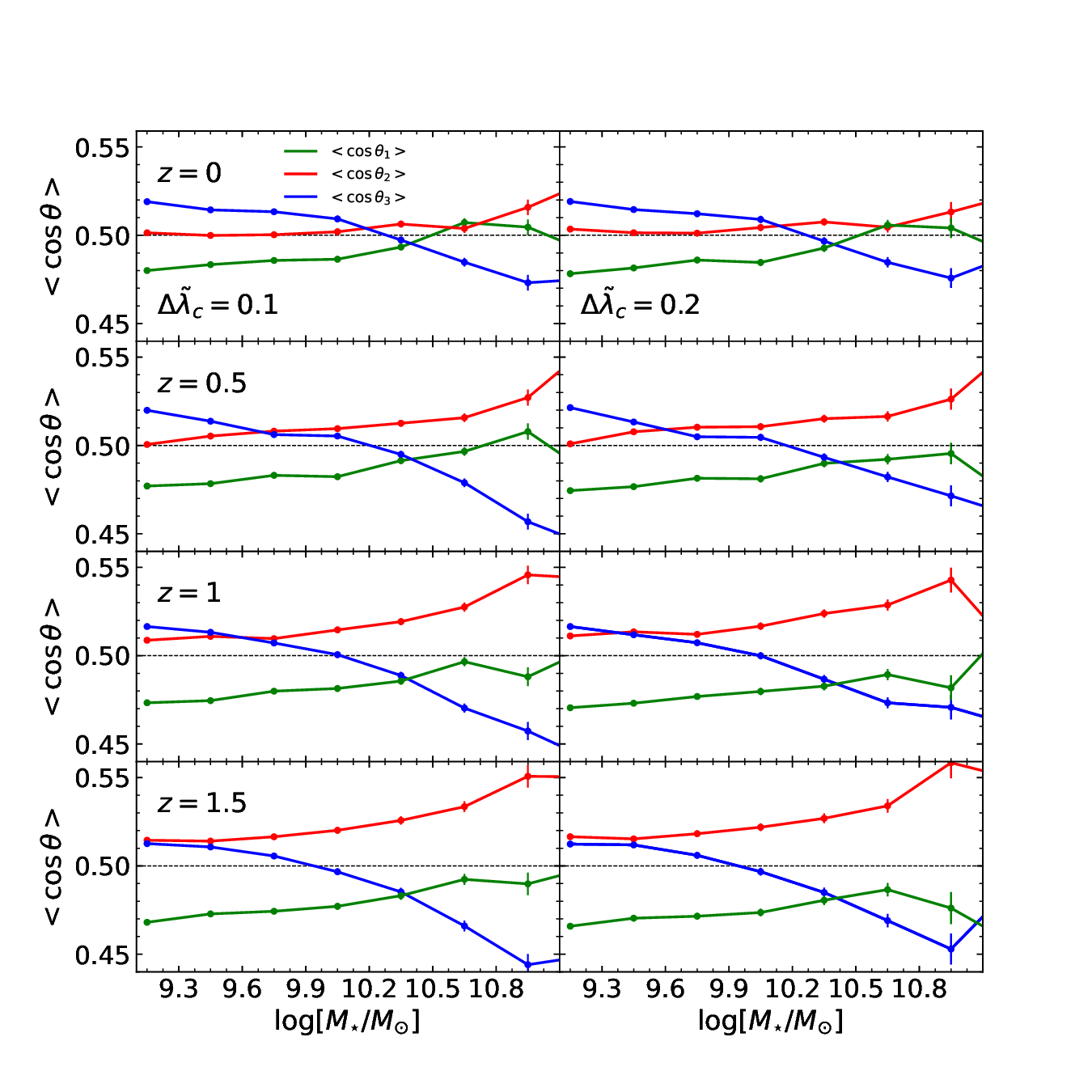}
\caption{Same as Figure \ref{fig:ez} but for two different cases of $\tlc$  with $R_{f}=2\,$Mpc at four different redshfits.}
\label{fig:ez_eig}
\end{center}
\end{figure}

\begin{figure}[ht]
\begin{center}
\plotone{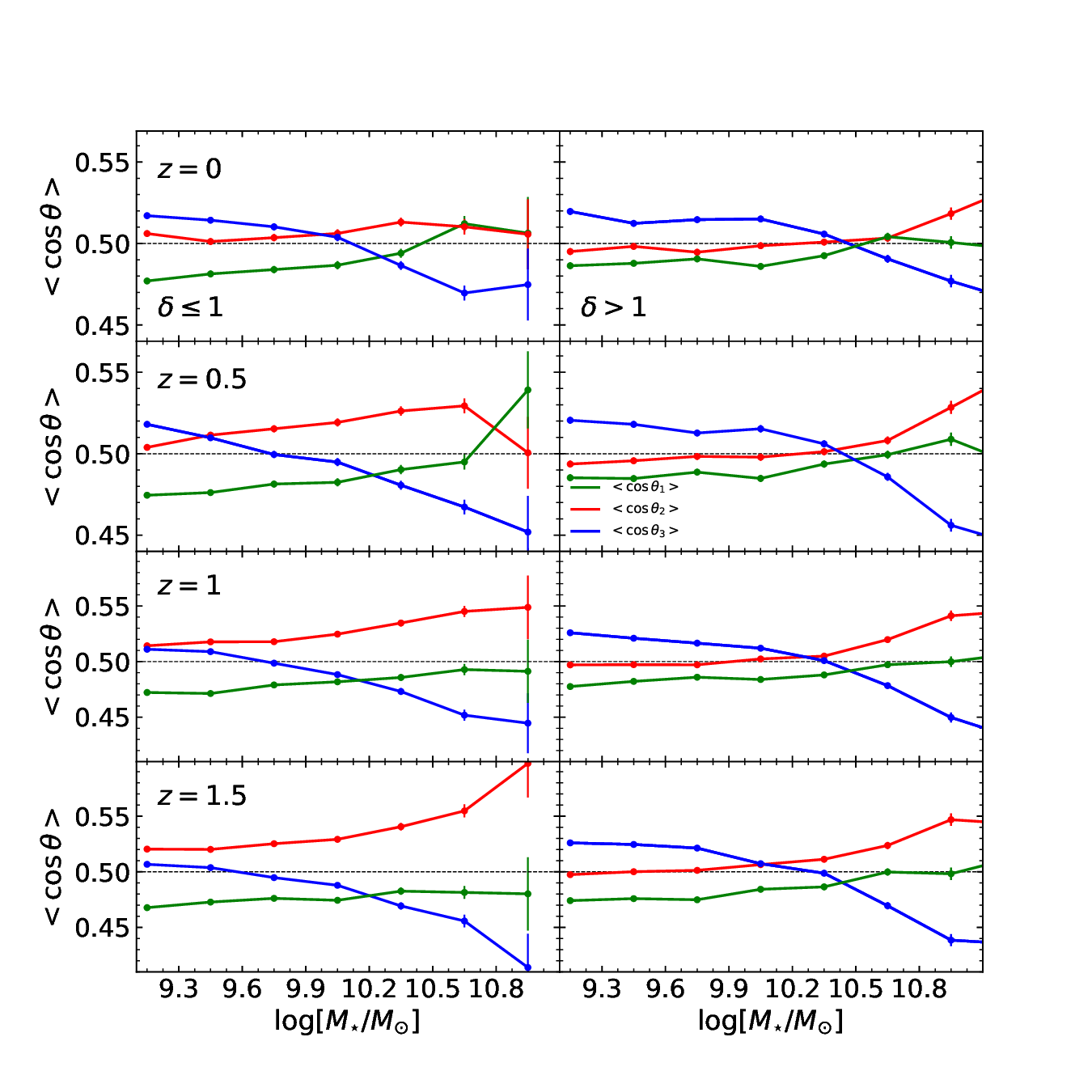}
\caption{Same as Figure \ref{fig:ez} but only for the galaxies located in the environments with low-density (left panels) 
and with high-density (right panels)  for the case of $R_{f}=2\,$Mpc.}
\label{fig:ez_den}
\end{center}
\end{figure}

\begin{figure}[ht]
\begin{center}
\plotone{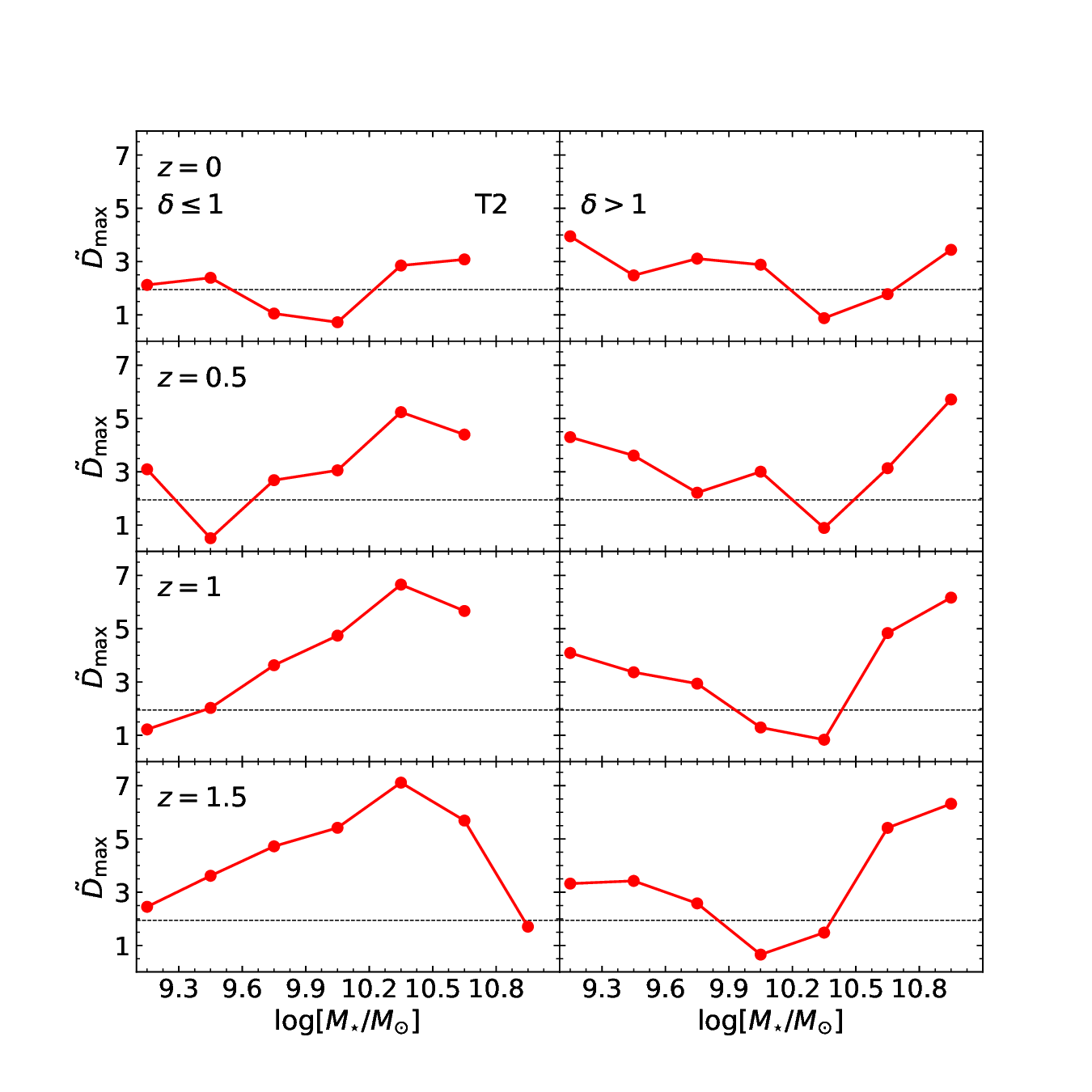}
\caption{KS statistics  for the determination of the T2 spin transition zones of the galaxies located in the low-density 
(left panels) and high-density (right panels) environments.}
\label{fig:clz_den}
\end{center}
\end{figure}

\begin{figure}[ht]
\begin{center}
\plotone{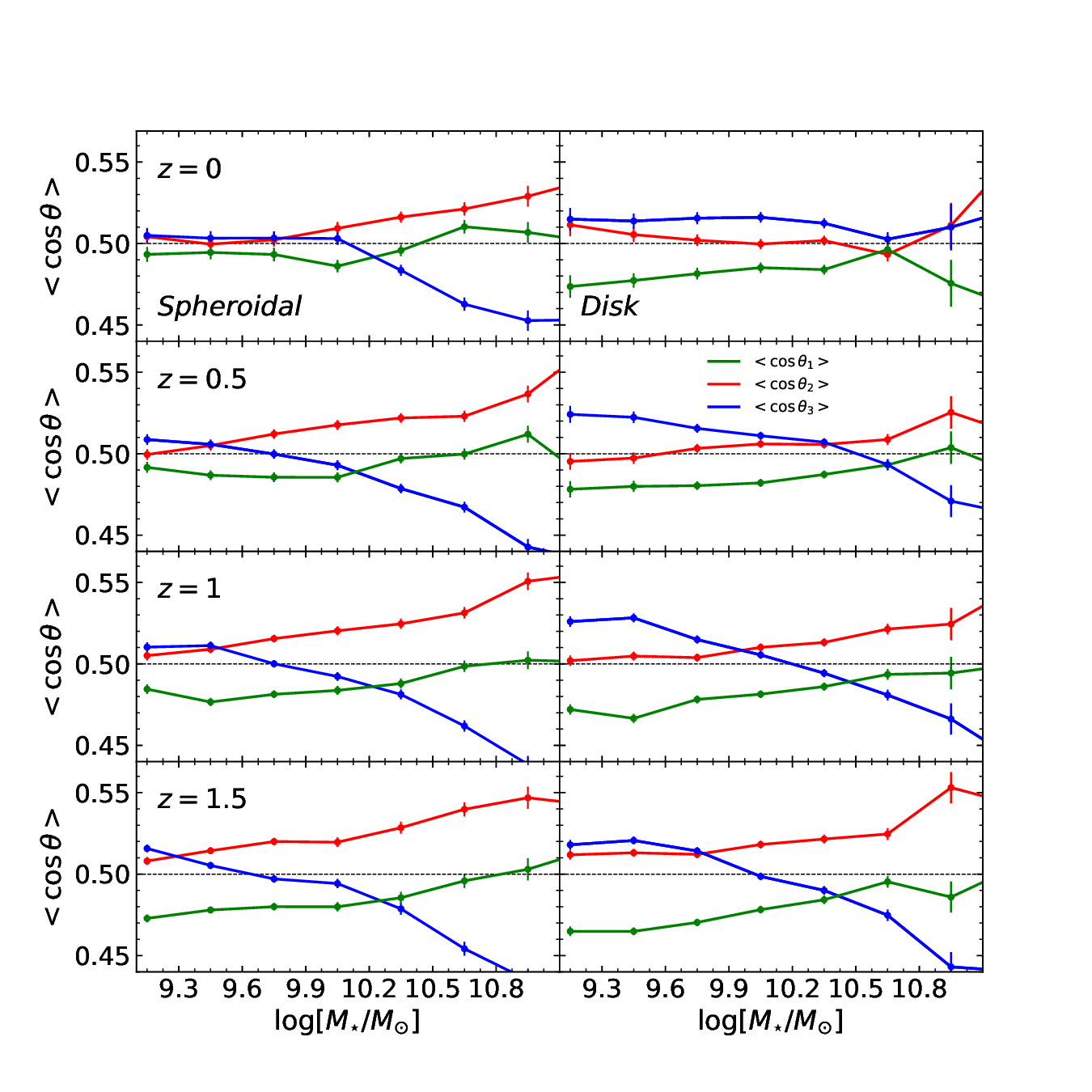}
\caption{Same as Figure \ref{fig:ez} but from the controlled samples of the spheroidals (left panels) and disks (right panels).}
\label{fig:ez_shape}
\end{center}
\end{figure}

\begin{figure}[ht]
\begin{center}
\plotone{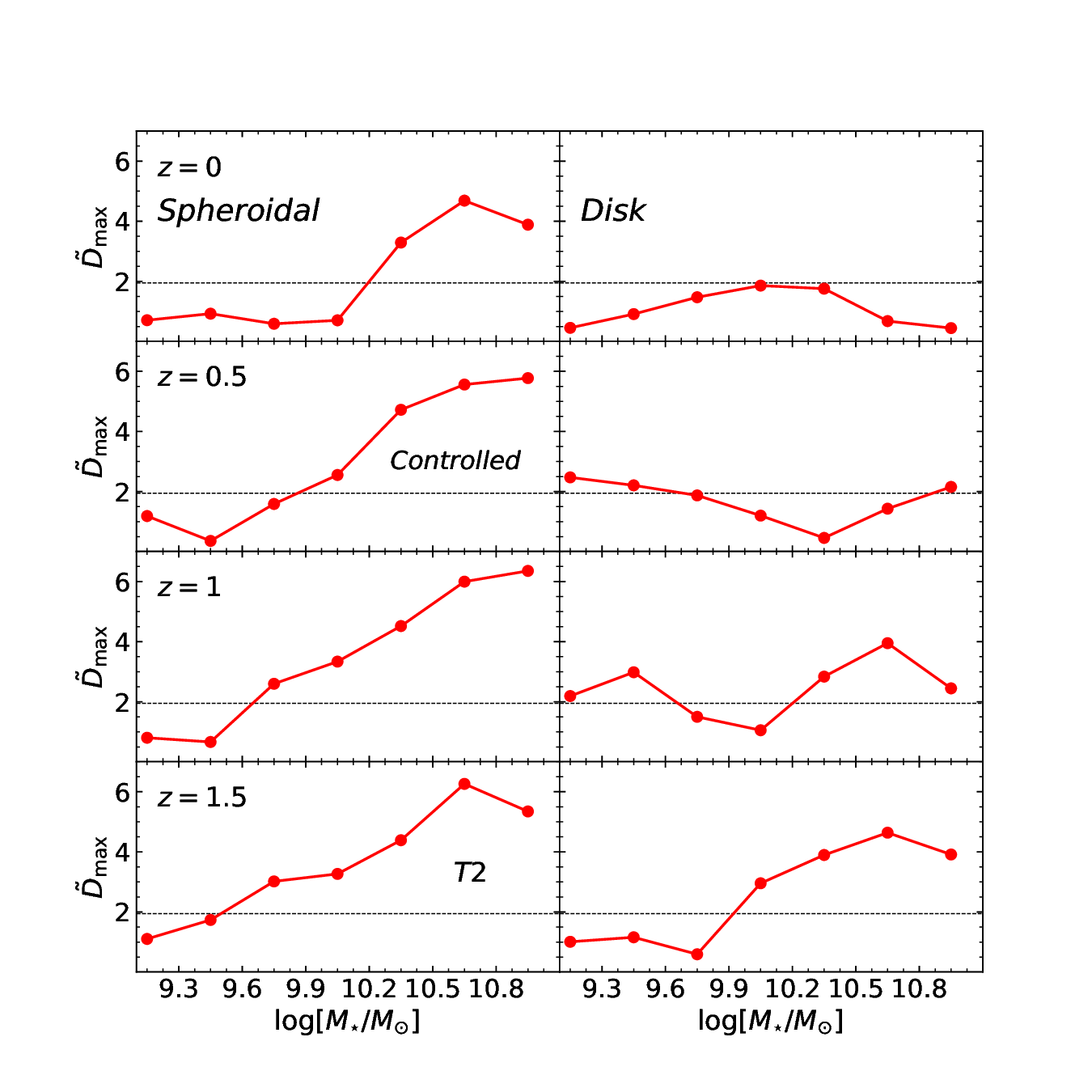}
\caption{KS statistics  for the determination of the T2 spin transition zones of the spheroidals (left panels) and of 
the disks (right panels) at  four different redshfits.}
\label{fig:clz_shape}
\end{center}
\end{figure}

\begin{figure}[ht]
\begin{center}
\plotone{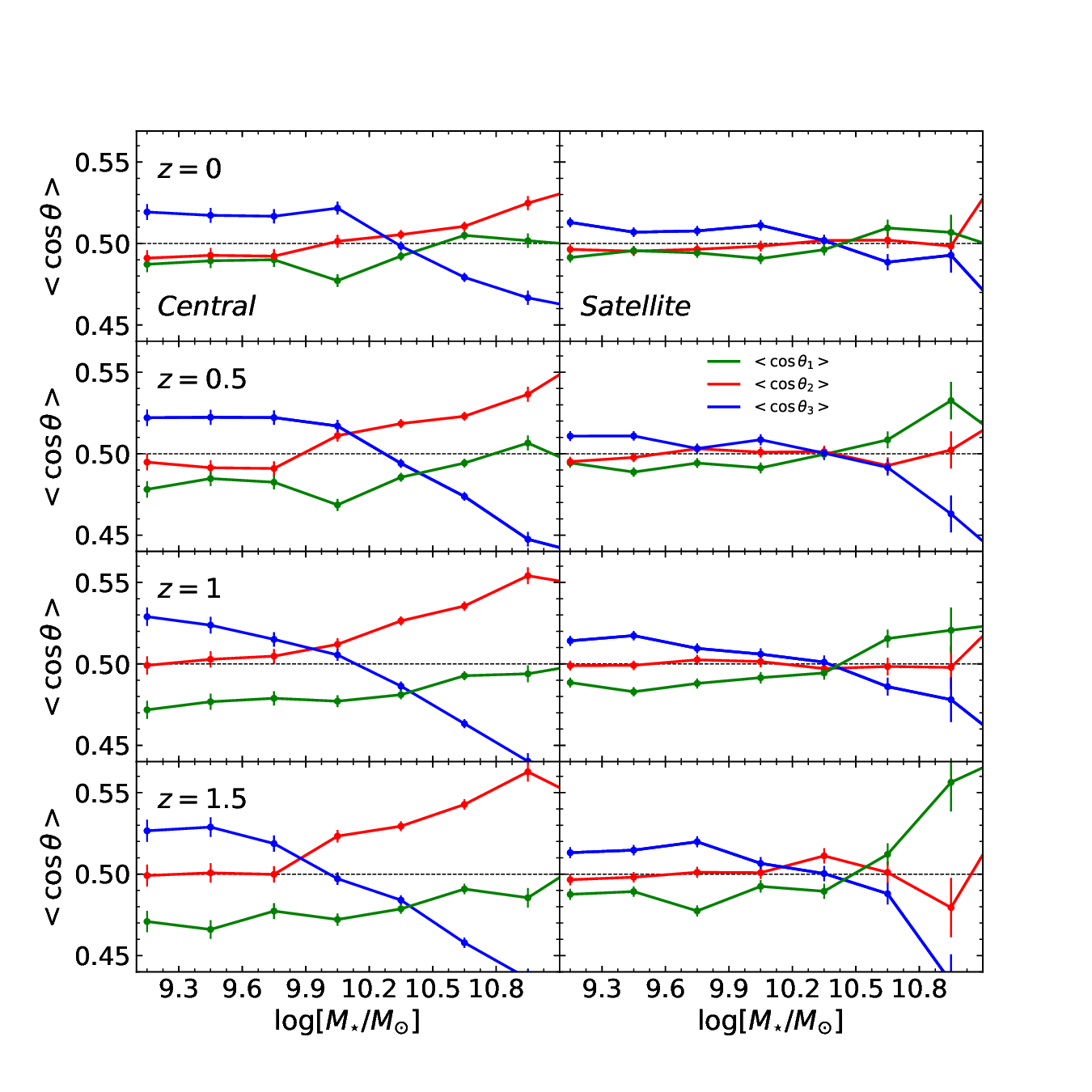}
\caption{Same as Figure \ref{fig:ez} but from the controlled samples of the centrals (left panels) and satellites (right panels) separately.}
\label{fig:ez_status}
\end{center}
\end{figure}

\begin{figure}[ht]
\begin{center}
\plotone{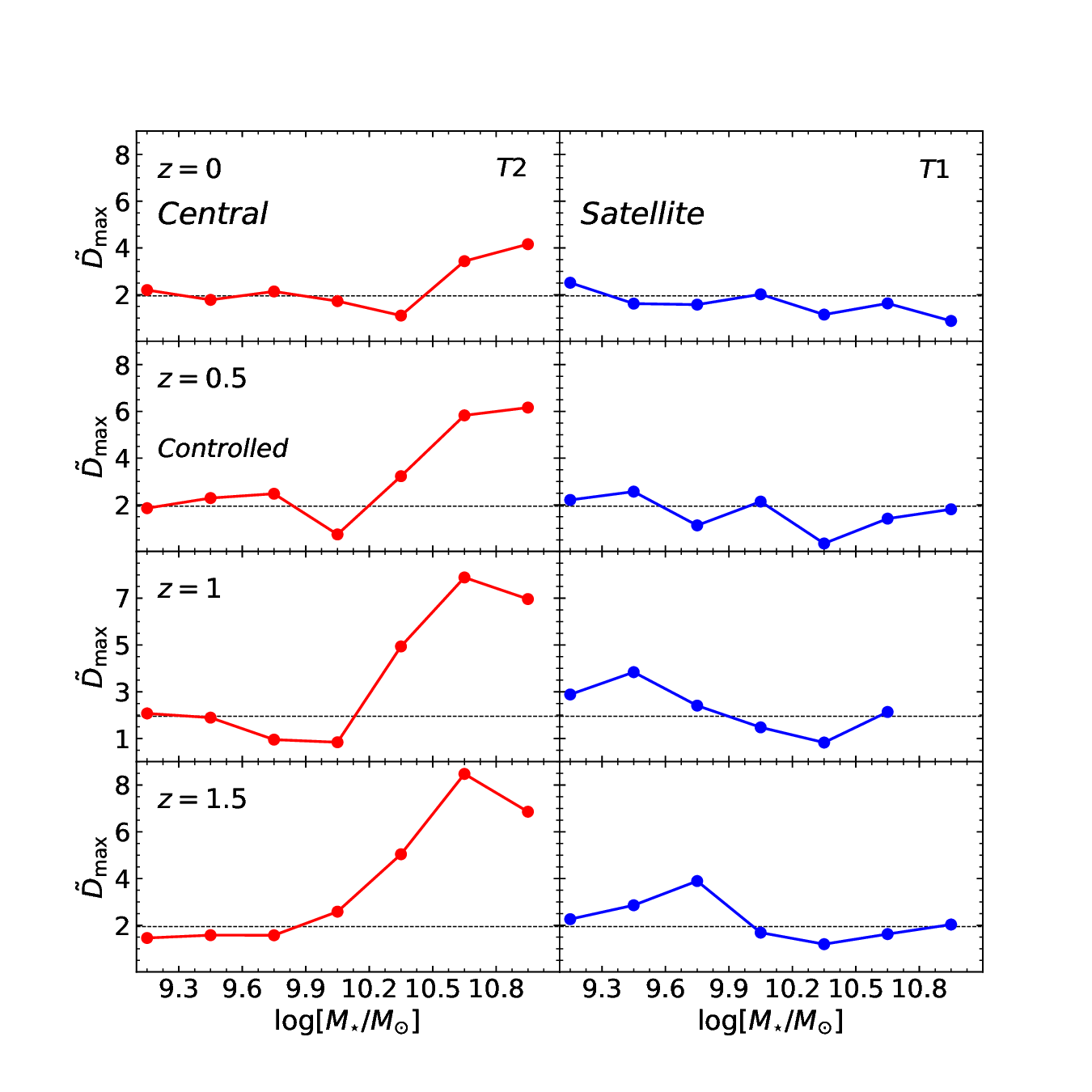}
\caption{KS statistics  for the determination of the T2 spin transition zones of the centrals (left panels) and 
 for the determination of the T1 spin transition zones of the satellites (right panels) at  four different redshfits.}
\label{fig:clz_status}
\end{center}
\end{figure}

\begin{figure}[ht]
\begin{center}
\plotone{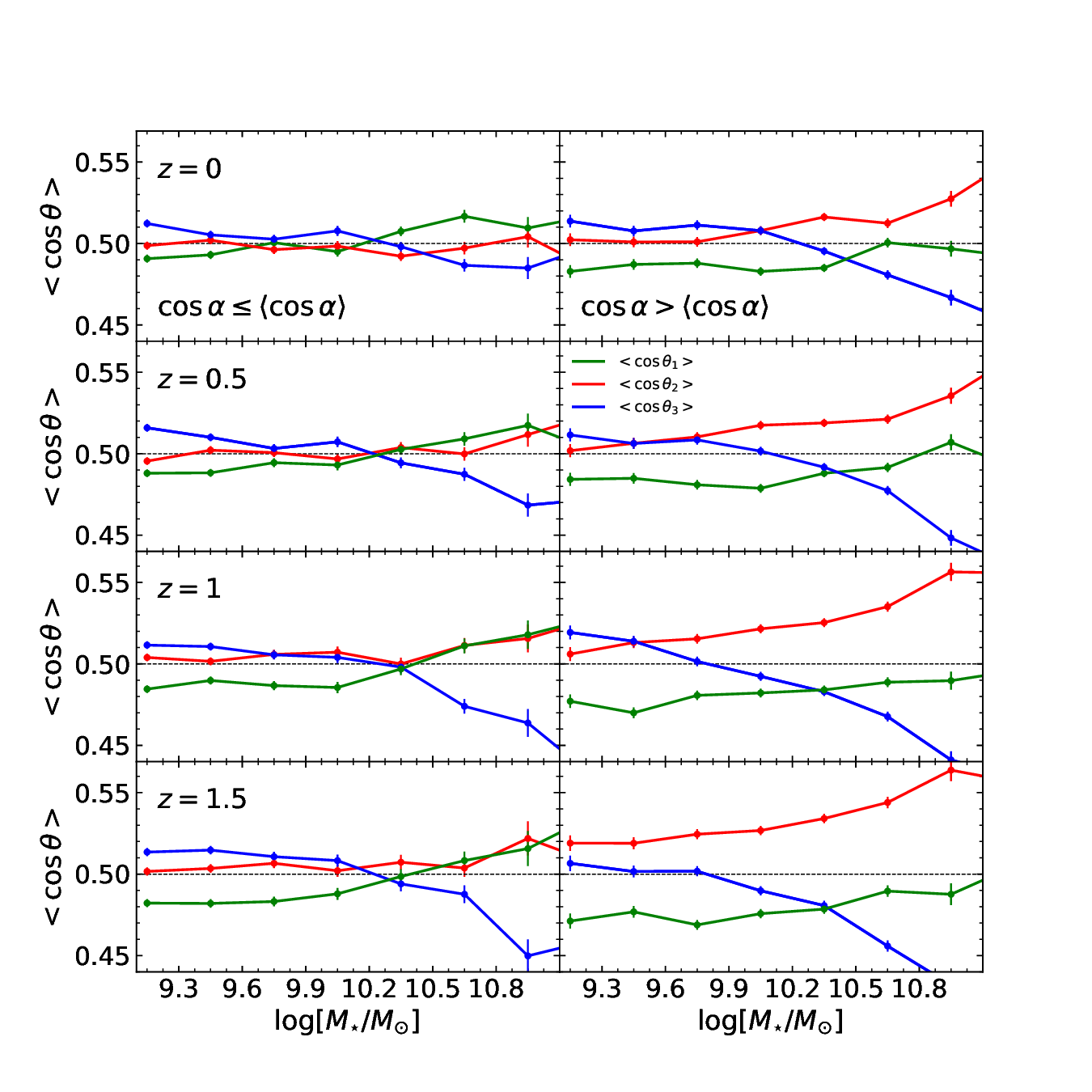}
\caption{Same as Figure \ref{fig:ez} but only for the galaxies with $\cos\alpha\le\langle\cos\alpha\rangle$ (left panels) and 
with $\cos\alpha>\langle\cos\alpha\rangle$ (right panels),  where $\alpha$ denotes the alignment angle between 
the baryon and total spins.}
\label{fig:ez_cosa}
\end{center}
\end{figure}

\begin{figure}[ht]
\begin{center}
\plotone{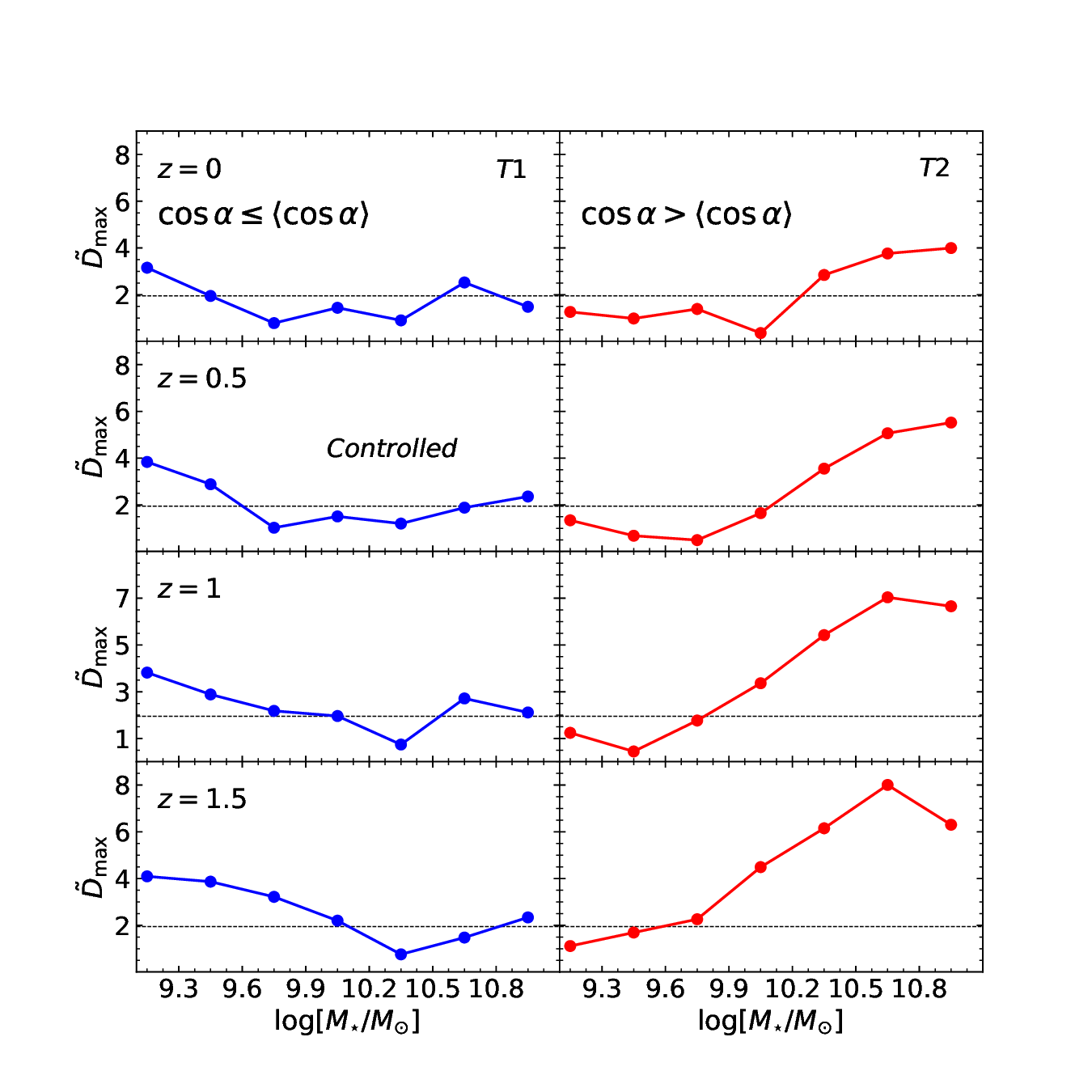}
\caption{KS statistics  for the determination of the T1 spin transition zones of the galaxies with 
$\cos\alpha\le\langle\cos\alpha\rangle$ (left panels) and  for the determination of the T2 spin transition zones 
of the galaxies with $\cos\alpha>\langle\cos\alpha\rangle$ (right panels) at  four different redshfits.}
\label{fig:clz_cosa}
\end{center}
\end{figure}

\begin{figure}[ht]
\begin{center}
\plotone{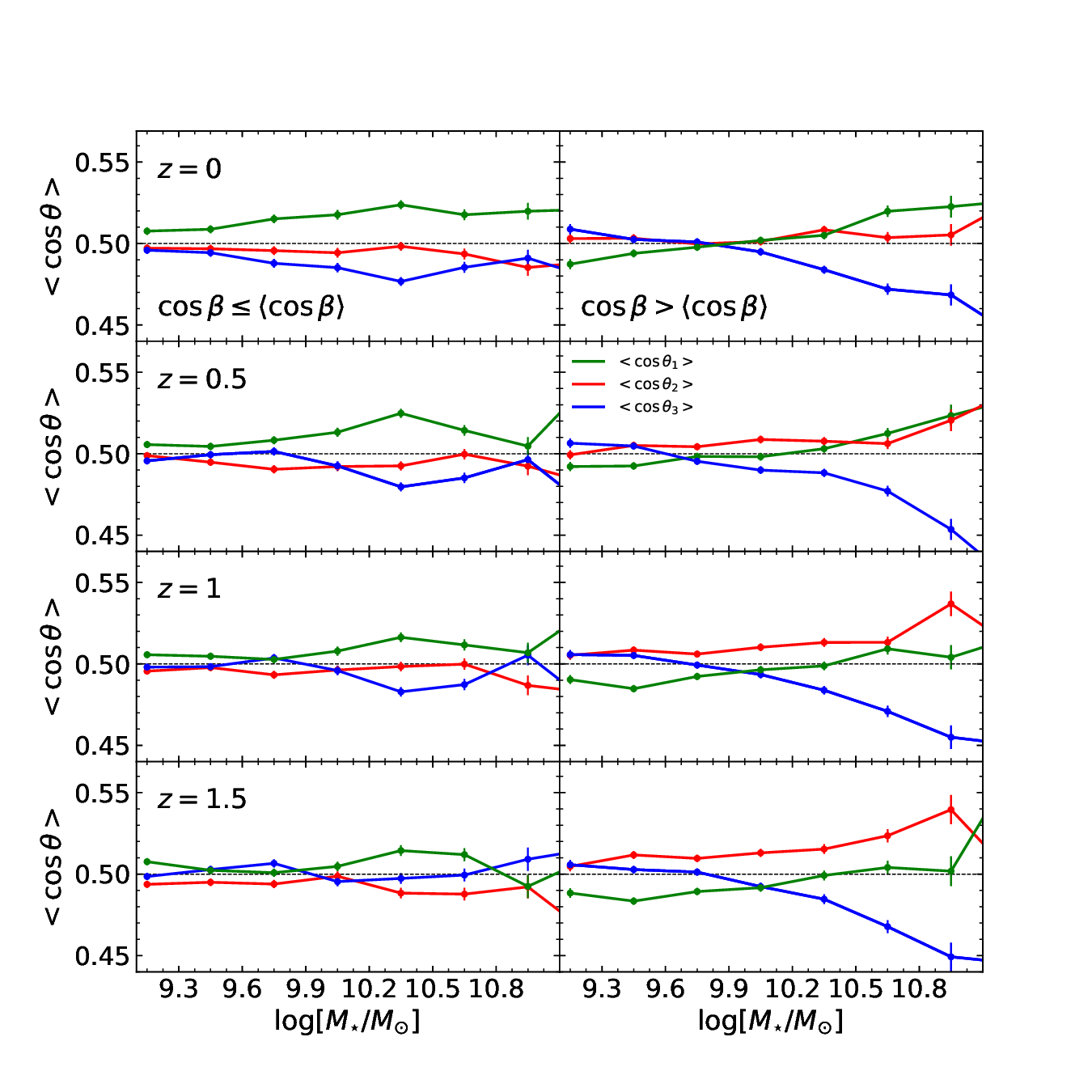}
\caption{Same as the middle panels of Figure \ref{fig:ez} but only for the galaxies with $\cos\beta\le\langle\cos\beta\rangle$ 
(left panels) and with $\cos\beta>\langle\cos\beta\rangle$ (right panels),  where $\beta$ denotes the alignment angle 
between the stellar and baryon spins.}
\label{fig:esz_cosb}
\end{center}
\end{figure}

\begin{figure}[ht]
\begin{center}
\plotone{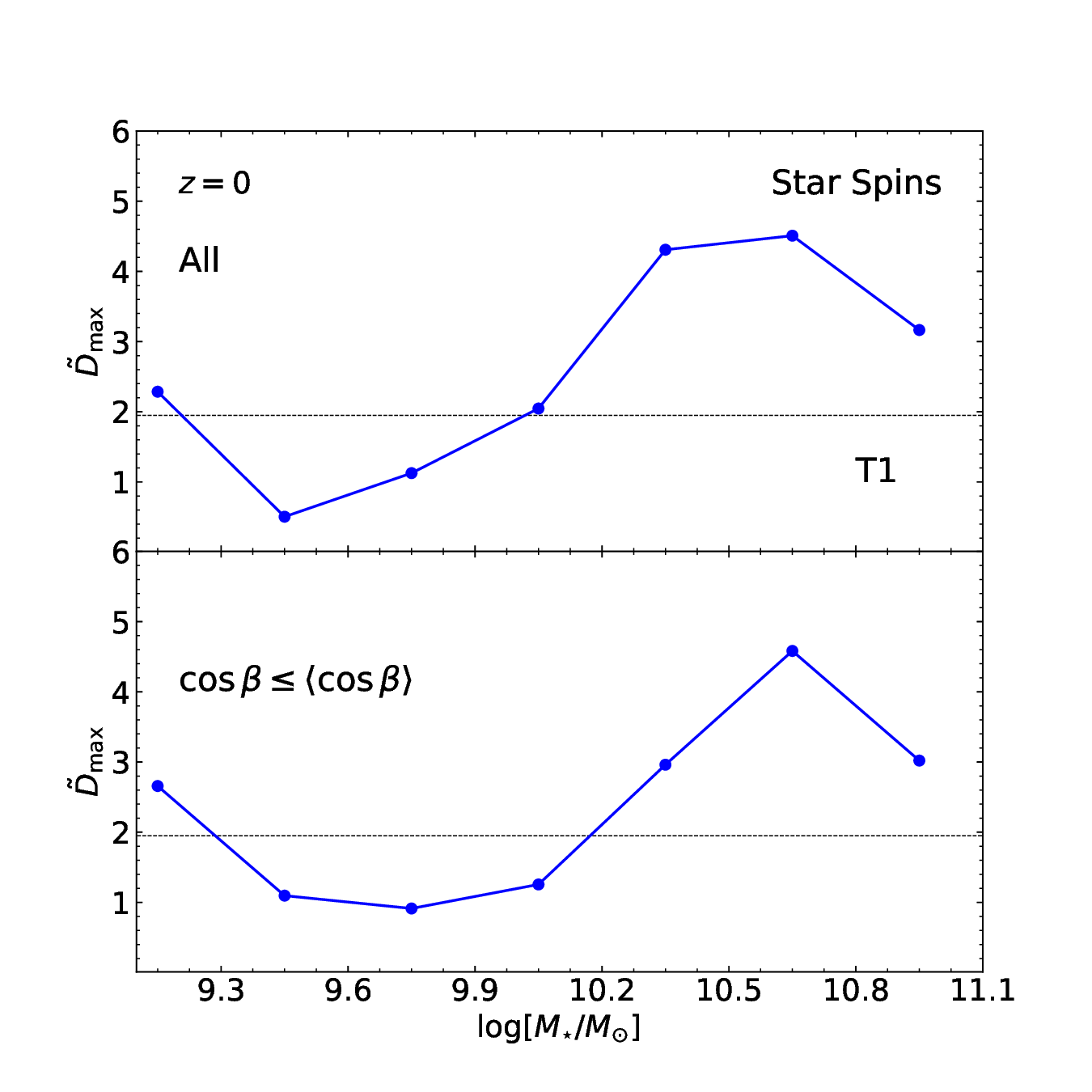}
\caption{KS statistics  for the determination of the T1 transition zones of the stellar spins of all galaxies 
(top panel) and of the galaxies with $\cos\beta\le\langle\cos\beta\rangle$ (bottom panel).}
\label{fig:clsz_cosb}
\end{center}
\end{figure}

The summary of our key results is the following. 
\begin{itemize}
\item 
The preferred directions of the galaxy baryon spins exhibit a mass dependent type-two (T2) transition from the minor to the intermediate tidal 
eigenvectors as $\ms$ increases. The T2 spin transition zones have been confirmed not to be sensitive to the eigenvalue degeneracy  but to 
be strongly dependent on $R_{f}$  in the redshift range of $0\le z\le 1$ (Figures \ref{fig:ez}-\ref{fig:ez_eig} and Tables \ref{tab:tzone1}-\ref{tab:tzone2}). 
\item
At $z=0$, the galaxies in the regions with $\delta>1$ undergo the T2 spin transitions in the higher mass ranges than the galaxies 
with $\delta\le 1$.  At $z=1$ and $1.5$ only the former yields significant signals of the T2 spin transition but in the lower mass ranges than 
at $z=0$ (Figures \ref{fig:ez_den}-\ref{fig:clz_den} and Table \ref{tab:tzone2}).  
\item
The disk galaxies undergo the T2 spin transition at $z=0.5$ and $1$, while no strong signal of spin transition is found from the spheroidal galaxies 
(Figures \ref{fig:ez_shape}-\ref{fig:clz_shape} and Table \ref{tab:tzone3}).  
\item
 The preferred directions of the baryon spins of the satellite galaxies exhibit a mass-dependent type-one (T1) transition from the minor to the major 
tidal eigenvectors as $\ms$ increases at $z=1$ and $1.5$. Whereas, the central galaxies exhibit the T2 spin transition at $z=0,\ 0.5$ and $1$
(Figures \ref{fig:ez_status}-\ref{fig:clz_status} and Table \ref{tab:tzone3}). 
\item
A clear signal of the T1 spin transition is found from the galaxies with  $\cos\alpha\le\langle\cos\alpha\rangle$ at all of the four redshifts, 
while no significant signal of the spin transition is found in the whole mass range of $9< \log(\ms/\munit)\le 11.1$ from the galaxies with 
$\cos\alpha>\langle\cos\alpha\rangle$ at any redshift (Figures \ref{fig:ez_cosa}-\ref{fig:clz_cosa} and Table \ref{tab:tzone3}). 
\item
The galaxy stellar spins yield a clear signal of the mass-dependent T1 transition at $z=0$  (top panel of Figure \ref{fig:clsz_cosb}). 
Significant signals of the mass-dependent T1 transition of the stellar spins are found only from those galaxies with 
$\cos\beta >\langle\cos\beta\rangle$ at $z=0$ (Figures \ref{fig:esz_cosb}-\ref{fig:clsz_cosb} and Table \ref{tab:tzone4}).
The stellar spins of the galaxies with $\cos\beta\le \langle\cos\beta\rangle$ are consistently aligned with the major tidal eigenvectors 
in the whole mass range at $z=0$ and $0.5$ (Figure \ref{fig:esz_cosb}). 
\end{itemize}

Our results are qualitatively inconsistent with the hydrodynamical results of \citet{cod-etal18} 
and \citet{ballet2}, who found no significant signals of the mass-dependent galaxy spin transitions at $z<0.5$, but consistent with those 
of \citet{wan-etal18} and \citet{kra-etal20} who reported detections of the significant signals even at $z=0$. 
Yet, it is difficult to make a parallel quantitative comparison between our results and those from the previous works because of the difference 
in the ways in which the spin alignments and transition mass ranges are measured. For example, in \citet{ballet2} and  \citet{kra-etal20} the galaxy 
stellar spins were measured relative only to the filament axes and their transition thresholds were determined by the criterion of 
$\langle\cost_{s,3}\rangle>0.5$. 

Given our result on the dependence of $\mts$ on $\delta$, we suspect that the $\bjb$-$\en$ alignments at $\ms< \mts$ are likely to be induced 
by some nonlinear evolutionary process, while the $\bjb$-$\ei$ alignments at $\ms> \mts$ should be induced by the tidal interactions in the linear and 
quasi-linear regimes, as predicted by the linear tidal torque theory \citep{LP00}. This logic can explain why only the galaxies in the high-density 
environments yield strong signals of the T2 spin transitions at $z>0$ as well as why their T2 spin transitions occur in the higher mass ranges at $z=0$ 
than the low-density counterparts.  In the high-density environments this nonlinear evolutionary mechanism should 
operate more actively even on the high mass scales at $z=0$ and even at higher redshifts.
 
The results on the morphology dependence of the T2 spin transition leads us to suspect that the spheroidal galaxies must be much less 
susceptible to this nonlinear process than the disks. Recalling the numerical finding of \citet{sal-etal12} that the strong spin alignments (misalignments) 
between the newly and previously accreted materials over time characterize the spheroidal (disk) galaxies,  this process responsible for the T2 spin transition 
might be more active in the regions where the coherent accretions of gas particles are restrained. 

Regarding our detection of the unique $\bjs$-$\ej$ alignments of the galaxy stellar spins and the T1 spin transition exhibited by the satellites,
we interpret it as a hint for the dependence of the galaxy spin alignment tendency on the gas temperature. 
If the gas cooling proceeds most efficiently in the plane normal to the major tidal eigenvectors, then the spin direction of the 
cold baryon gas would acquire the $\bjs$-$\ej$ alignment tendency, while the hot baryon spins still exhibit the same $\bjs$-$\ei$ alignment 
tendency as the total or DM spins. The T1 spin transition exhibited by the satellite galaxies may be explained if those satellites with $\ms>\mtp$ 
contain higher fraction of cold gas than those with $\ms\le\mtp$.

These scenarios, however, are mere speculations with no backup numerical evidences. To verify our speculations and to figure out what exactly these 
mechanisms are, it will be necessary to conduct a much thorough systematic work in which the trajectories of all  baryonic particles  
involved in the merging history of their host subhalos can be traced. 
Our future work will be in the direction of performing such a comprehensive numerical analysis as well as finding an observational evidence for the 
occurrence of the mass dependent {\it dual-type} spin transition of real galaxies. 

\acknowledgments

The IllustrisTNG simulations were undertaken with compute time awarded by the Gauss Centre for Supercomputing (GCS) under GCS Large-Scale Projects GCS-ILLU 
and GCS-DWAR on the GCS share of the supercomputer Hazel Hen at the High Performance Computing Center Stuttgart (HLRS), as well as on the machines of the Max 
Planck Computing and Data Facility (MPCDF) in Garching, Germany.

We are very grateful to an anonymous referee whose scrupulous review and constructive criticisms helped us significantly improve the original manuscript. 
JL thank K.Kraljic for useful comments.
JL and SR acknowledges the support by Basic Science Research Program through the National Research Foundation (NRF) of Korea 
funded by the Ministry of Education (No.2019R1A2C1083855). S.-J.Y. and J.-S. M. acknowledge support from the Mid-career
Researcher Program (No. 2019R1A2C3006242) through the NRF of Korea.

\clearpage
%%%%%%%%%%%%%%%%%%%%%%%%%%%%%%%%%%%%%%%%%%%%%%%%%%%%%%%
%%%%%%%%%%%%%%%%%%%%%%%%%%%%%%%%%%%%%%%%%%%%%%%%%%%%%%%%
\clearpage
%%%%%%%%%%%%%%%%%%%%%%%%%%%%%%%%%%%%%%%%%%%%%%%%%%%%%%%
%%%%%%%%%%%%%%%%%%%%%%%%%%%%%%%%%%%%%%%%%%%%%%%%%%%%%%%%
\clearpage
%%%%%%%%%%%%%%%%%%%%%%%%%%%%%%%%%%%%%%%%%%%%%%%%%%%%%%%
%%%%%%%%%%%%%%%%%%%%%%%%%%%%%%%%%%%%%%%%%%%%%%%%%%%%%%%%
\clearpage
%%%%%%%%%%%%%%%%%%%%%%%%%%%%%%%%%%%%%%%%%%%%%%%%%%%%%%%
%%%%%%%%%%%%%%%%%%%%%%%%%%%%%%%%%%%%%%%%%%%%%%%%%%%%%%%%
\clearpage
%%%%%%%%%%%%%%%%%%%%%%%%%%%%%%%%%%%%%%%%%%%%%%%%%%%%%%%
%%%%%%%%%%%%%%%%%%%%%%%%%%%%%%%%%%%%%%%%%%%%%%%%%%%%%%%%
\clearpage
%%%%%%%%%%%%%%%%%%%%%%%%%%%%%%%%%%%%%%%%%%%%%%%%%%%%%%%
%%%%%%%%%%%%%%%%%%%%%%%%%%%%%%%%%%%%%%%%%%%%%%%%%%%%%%%%
\clearpage
%%%%%%%%%%%%%%%%%%%%%%%%%%%%%%%%%%%%%%%%%%%%%%%%%%%%%%%
%%%%%%%%%%%%%%%%%%%%%%%%%%%%%%%%%%%%%%%%%%%%%%%%%%%%%%%%
\clearpage
%%%%%%%%%%%%%%%%%%%%%%%%%%%%%%%%%%%%%%%%%%%%%%%%%%%%%%%
%%%%%%%%%%%%%%%%%%%%%%%%%%%%%%%%%%%%%%%%%%%%%%%%%%%%%%%%
\clearpage
%%%%%%%%%%%%%%%%%%%%%%%%%%%%%%%%%%%%%%%%%%%%%%%%%%%%%%%
%%%%%%%%%%%%%%%%%%%%%%%%%%%%%%%%%%%%%%%%%%%%%%%%%%%%%%%%
\clearpage
%%%%%%%%%%%%%%%%%%%%%%%%%%%%%%%%%%%%%%%%%%%%%%%%%%%%%%%
%%%%%%%%%%%%%%%%%%%%%%%%%%%%%%%%%%%%%%%%%%%%%%%%%%%%%%%%
\clearpage
%%%%%%%%%%%%%%%%%%%%%%%%%%%%%%%%%%%%%%%%%%%%%%%%%%%%%%%
%%%%%%%%%%%%%%%%%%%%%%%%%%%%%%%%%%%%%%%%%%%%%%%%%%%%%%%%
\clearpage
%%%%%%%%%%%%%%%%%%%%%%%%%%%%%%%%%%%%%%%%%%%%%%%%%%%%%%%
%%%%%%%%%%%%%%%%%%%%%%%%%%%%%%%%%%%%%%%%%%%%%%%%%%%%%%%%
\clearpage
%%%%%%%%%%%%%%%%%%%%%%%%%%%%%%%%%%%%%%%%%%%%%%%%%%%%%%%
%%%%%%%%%%%%%%%%%%%%%%%%%%%%%%%%%%%%%%%%%%%%%%%%%%%%%%%%
\clearpage
%%%%%%%%%%%%%%%%%%%%%%%%%%%%%%%%%%%%%%%%%%%%%%%%%%%%%%%
%%%%%%%%%%%%%%%%%%%%%%%%%%%%%%%%%%%%%%%%%%%%%%%%%%%%%%%%
\clearpage
%%%%%%%%%%%%%%%%%%%%%%%%%%%%%%%%%%%%%%%%%%%%%%%%%%%%%%%
%%%%%%%%%%%%%%%%%%%%%%%%%%%%%%%%%%%%%%%%%%%%%%%%%%%%%%%%
\clearpage
%%%%%%%%%%%%%%%%%%%%%%%%%%%%%%%%%%%%%%%%%%%%%%%%%%%%%%%
%%%%%%%%%%%%%%%%%%%%%%%%%%%%%%%%%%%%%%%%%%%%%%%%%%%%%%%%
\clearpage
\begin{deluxetable}{ccc}
\tablewidth{0pt}
\setlength{\tabcolsep}{3mm}
\tablecaption{Numbers of the Galaxies}
\tablehead{redshift & $N_{\rm fof}$ & $N_{g}$}
\startdata
$0$ & $17625892$ & $218943$    \\
$0.5$ & $18725103$ & $215319$  \\
$1$ & $19595267$ & $202521$  \\
$1.5$ & $20332985$ & $174280$  \\
\enddata
\label{tab:ng}
\end{deluxetable}
%%%%%%%%%%%%%%%%%%%%%%%%%%%%%%%%%%%%%%%%%%%%%%%%%%%%%%%%
%%%%%%%%%%%%%%%%%%%%%%%%%%%%%%%%%%%%%%%%%%%%%%%%%%%%%%%%
\clearpage
\begin{deluxetable}{ccc}
\tablewidth{0pt}
\setlength{\tabcolsep}{3mm}
\tablecaption{T2 Transition Zones of the Baryon Spins}
\tablehead{redshift & $R_{f}$ & $\log(\mts/\munit)$  \\  & (Mpc) &  }
\startdata
$0$ & $2$ & $(9.9,\ 10.5)$   \\
$0.5$ & $2$ & $(9.6\, ,\ 10.2)$  \\
$1$ & $2$ & $(9.3\, ,\ 9.9)$  \\
$1.5$ & $2$ & --  \\
\hline
$0$ & $3$ & $(10.2,\ 10.5)$   \\
$0.5$ & $3$ & $(9.6\, ,\ 10.2)$  \\
$1$ & $3$ & $(9.6\, ,\ 10.2)$  \\
$1.5$ & $3$ & --  \\
\hline
$0$ & $5$ & $(10.2,\ 10.8)$   \\
$0.5$ & $5$ & $(9.6\, ,\ 10.5)$  \\
$1$ & $5$ & $(9.6,\ 10.5)$  \\
$1.5$ & $5$ & $(9.3\, ,\ 10.2)$  \\
\enddata
\label{tab:tzone1}
\end{deluxetable}
%%%%%%%%%%%%%%%%%%%%%%%%%%%%%%%%%%%%%%%%%%%%%%%%%%%%%%%%
%%%%%%%%%%%%%%%%%%%%%%%%%%%%%%%%%%%%%%%%%%%%%%%%%%%%%%%%
\clearpage
\begin{deluxetable}{cccccc}
\tablewidth{0pt}
\setlength{\tabcolsep}{3mm}
\tablecaption{Effects of the Eigenvalue Degeneracy and Environmental Density on $\mts$}
\tablehead{redshift & $N_{g}$ & $R_{f}$ & $\Delta\tilde{\lambda}_{c}$ & $\delta$ & $\log(\mts/\munit)$  
\\  & & (Mpc) & & &}
\startdata
$0$ & $177323$ & $2$ & $0.1$ & All & $(9.9\, ,\ 10.5)$   \\
$0.5$ & $184274$ & $2$ & $0.1$ & All & $(9.9\, ,\ 10.2)$   \\
$1$ & $178886$ & $2$ & $0.1$ & All & $(9.3\, ,\ 9.9)$  \\
$1.5$ & $156088$ & $2$ & $0.1$ & All & --  \\
\hline
$0$ & $136664$ & $2$ & $0.2$ & All & $(9.6\, ,\ 10.5)$   \\
$0.5$ & $143000$ & $2$ & $0.2$ & All & $(9.3\, ,\ 10.2)$   \\
$1$ & $139996$ & $2$ & $0.2$ & All & --  \\
$1.5$ & $122019$ & $2$ & $0.2$ & All & --  \\
\hline
$0$ & $127066$ & $2$ & $0$ & $> 1$ & $(10.2\, ,\ 10.8)$   \\
$0.5$ & $115803$ & $2$ & $0$ & $> 1$ & $(10.2\, ,\ 10.5)$   \\
$1$ & $98120$ & $2$ & $0$ & $> 1$ & $(9.9\, ,\ 10.5)$ \\
$1.5$ & $74629$ & $2$ & $0$ & $> 1$ & $(9.9\, ,\ 10.5)$  \\
\hline
$0$ & $91877$ & $2$ & $0$ & $\le 1$ & $(9.6\, ,\ 10.2)$   \\
$0.5$ & $99516$ & $2$ & $0$ & $\le 1$ & $(9.3\, ,\ 9.6)$   \\
$1$ & $104401$ & $2$ & $0$ & $\le 1$ & -- \\
$1.5$ & $99651$ & $2$ & $0$ & $\le 1$ & --  \\
\enddata
\label{tab:tzone2}
\end{deluxetable}
%%%%%%%%%%%%%%%%%%%%%%%%%%%%%%%%%%%%%%%%%%%%%%%%%%%%%%%%
\clearpage
\begin{deluxetable}{ccccccc}
\tablewidth{0pt}
\setlength{\tabcolsep}{3mm}
\tablecaption{Dependence of the Spin Transition Zones on the Galaxy Properties}
\tablehead{redshift & $N_{g}$ & Status & Morphology & $\cos\alpha$ & $\log(\mts/\munit)$  & $\log(\mtp/\munit)$}
\startdata
$0.5$ & $59291$ & All & Disk & All & $(9.9\, ,\ 10.8)$ &  --  \\
$1$ & $71526$ & All & Disk & All & $(9.6\, ,\ 10.2)$ & --  \\
\hline
$0$ & $46978$ & Central & All & All & $(9.9\, ,\ 10.5)$ & --  \\
$0.5$ & $48017$ & Central & All & All & $(9.9\, ,\ 10.2)$ & --  \\
$1$ & $43866$ & Central & All & All & $(9.6\, ,\ 10.2)$ & --  \\
$1$ & $43866$ & Satellite & All & All & -- & $(9.9\, ,\ 10.5)$ \\
$1.5$ & $35042$ & Satellite & All & All & -- & $(9.9\, ,\ 10.8)$ \\
\hline
$0$ & $66706$ & All & All & $\le 0.826$ & -- & $(9.6\, ,\ 10.5)$ \\
$0.5$ & $65138$ & All & All & $\le 0.831$ & -- & $(9.6\, ,\ 10.5)$ \\
$1$ & $59856$ & All & All & $\le 0.841$ & -- & $(10.2\, ,\ 10.5)$  \\
$1.5$ & $50319$ & All & All & $\le 0.854$ & -- & $(10.2\, ,\ 10.8)$  \\
\enddata
\label{tab:tzone3}
\end{deluxetable}
%%%%%%%%%%%%%%%%%%%%%%%%%%%%%%%%%%%%%%%%%%%%%%%%%%%%%%%%
%%%%%%%%%%%%%%%%%%%%%%%%%%%%%%%%%%%%%%%%%%%%%%%%%%%%%%%%
\clearpage
\begin{deluxetable}{cccc}
\tablewidth{0pt}
\setlength{\tabcolsep}{3mm}
\tablecaption{T1 Transition Zones of the Stellar Spins}
\tablehead{redshift & $N_{g}$ & $\cos\beta$ & $\log(\mtp/\munit)$ }
\startdata
$0$ & $218943$ & All & $(9.3,\ 9.9)$   \\
$0$ & $77464$ & $> 0.756$ &  $(9.3,\ 10.2)$  \\
\enddata
\label{tab:tzone4}
\end{deluxetable}
%%%%%%%%%%%%%%%%%%%%%%%%%%%%%%%%%%%%%%%%%%%%%%%%%%%%%%%%
%%%%%%%%%%%%%%%%%%%%%%%%%%%%%%%%%%%%%%%%%%%%%%%%%%%%%%%%
\end{document}